\documentclass{article}
 
\usepackage[a4paper,left=20mm,top=25mm,right=20mm,bottom=35mm]{geometry}

\usepackage{authblk}
\usepackage[numbers, sort&compress]{natbib}
\usepackage[margin=30pt,font=small,labelfont=bf]{caption}
\usepackage{amsmath,amsopn,amsthm,amssymb}

\usepackage{graphicx}
\usepackage{mathptmx} 
\usepackage[mathscr]{euscript}
\usepackage{mathtools}
\usepackage{float}
\usepackage{enumitem}
\usepackage{xspace}
\usepackage{scalerel}

\usepackage[
 	colorlinks=true,
	urlcolor=black,
	linkcolor=blue
]{hyperref}

\usepackage{algorithm}
\usepackage[noend]{algpseudocode}
\algrenewcommand\algorithmicrequire{\textbf{Input:}}
\algrenewcommand\algorithmicensure{\textbf{Output:}}

\usepackage{tabularx}
\makeatletter
\newcommand{\multiline}[1]{%
  \begin{tabularx}{\dimexpr\linewidth-\ALG@thistlm}[t]{@{}X@{}}
    #1
  \end{tabularx}
}
\makeatother

\usepackage{algorithmicx}
\usepackage{algorithm} 
\usepackage[noend]{algpseudocode}
\algrenewcommand\algorithmicrequire{\textbf{Input:}}
\algrenewcommand\algorithmicensure{\textbf{Output:}}

\newtheorem{theorem}{Theorem}[section]

\newtheorem{lemma}[theorem]{Lemma} 
\newtheorem{corollary}[theorem]{Corollary}
\newtheorem{proposition}[theorem]{Proposition}

\newtheorem{definition}[theorem]{Definition}
\newtheorem{observation}[theorem]{Observation}

\newtheorem*{remark}{Remark}

\makeatletter
\def\moverlay{\mathpalette\mov@rlay}
\def\mov@rlay#1#2{\leavevmode\vtop{%
   \baselineskip\z@skip \lineskiplimit-\maxdimen
   \ialign{\hfil$\m@th#1##$\hfil\cr#2\crcr}}}
\newcommand{\charfusion}[3][\mathord]{
    #1{\ifx#1\mathop\vphantom{#2}\fi
        \mathpalette\mov@rlay{#2\cr#3}
      }
    \ifx#1\mathop\expandafter\displaylimits\fi}
\makeatother

\newcommand{\cupdot}{\charfusion[\mathbin]{\cup}{\cdot}}

\DeclareMathOperator{\union}{\cupdot}
\DeclareMathOperator{\join}{\mathrel{\ooalign{\hss$\triangleleft$\hss\cr$\triangleright$}}}

\def\CircleArrowleft{\ensuremath{%
  \reflectbox{\rotatebox[origin=c]{180}{$\circlearrowleft$}}}}
\DeclareMathOperator{\inflateOP}{\CircleArrowleft}

\newcommand{\Hasse}[1][]{\mathscr{H}\ifthenelse{\equal{#1}{}}{}{(#1)}}
\newcommand{\hasse}[1][]{\mathscr{H}\ifthenelse{\equal{#1}{}}{}{(#1)}}
\DeclareMathOperator{\CC}{\mathtt{C}}

\newcommand{\levIex}[1][1]{\upshape{\textsc{Lev-{#1}-Ex}}\xspace}

\newcommand{\lca}{\ensuremath{\operatorname{lca}}}
\newcommand{\LCA}{\ensuremath{\operatorname{LCA}}}
\newcommand{\NWA}{\scaleto{\nwarrow}{4pt}}
\newcommand{\atomE}{\mathsf{T}}
\newcommand{\MD}{\ensuremath{\mathbb{M}}}
\newcommand{\MDstrong}{\ensuremath{\mathbb{M}_{\mathrm{str}}}}
\newcommand{\Mmax}{\ensuremath{\mathbb{M}_{\max}}}
\newcommand{\MDT}{\ensuremath{\mathscr{T}}}
\DeclareMathOperator{\pvr}{pvr}
\newcommand{\pfam}{\mathcal{P}}
\renewcommand{\P}{\ensuremath{\mathbb{P}}}
\DeclareMathOperator{\child}{child}
\DeclareMathOperator{\parent}{parent}
\DeclareMathOperator{\indeg}{indeg}
\DeclareMathOperator{\outdeg}{outdeg}

\DeclareMathOperator{\LL}{L}
\DeclareMathOperator{\expand}{\mathrm{EXPD}}

\newcommand{\primeL}{\mathrm{prime}}
\DeclareMathOperator{\tww}{tww}

    {\begin{description}[leftmargin = 0.2cm, labelsep = 0.2cm]}
    {\end{description}}
\providecommand{\keywords}[1]{\textbf{\textit{Keywords: }} #1}

\title{Orthology and Near-Cographs in the Context of Phylogenetic Networks}

\author[1]{Anna Lindeberg} 

\author[2]{Guillaume E. Scholz} 

\author[3]{Nicolas Wieseke} 

\author[1,*]{Marc Hellmuth} 

\affil[1]{Department of Mathematics, Faculty of Science,
  Stockholm University, SE-10691 Stockholm, Sweden} 

\affil[2]{Bioinformatics Group, Department of Computer Science \&
    Interdisciplinary Center for Bioinformatics, Universität Leipzig,
    Härtelstra{\ss}e~16--18, D-04107 Leipzig, Germany.}

\affil[3]{Swarm Intelligence and Complex Systems Group, Department of Computer Science,
    Universität Leipzig, Augustusplatz 10, D-04109 Leipzig, Germany.}

\affil[*]{corresponding author}

\date{\ }

\begin{document}
\sloppy

\maketitle

\abstract{ 

Orthologous genes, which arise through speciation, play a key role in comparative genomics and
functional inference. In particular, graph-based methods allow for the inference of orthology
estimates without prior knowledge of the underlying gene or species trees. This results in
orthology graphs, where each vertex represents a gene, and an edge exists between two vertices if
the corresponding genes are estimated to be orthologs.
Orthology graphs inferred under a tree-like evolutionary model must be cographs.
However, real-world
data often deviate from this property, either due to noise in the data, errors in inference methods or, 
simply, because evolution follows a network-like rather than a tree-like process. The latter, in particular, raises the 
question of whether and how orthology graphs can be derived from
or, equivalently, are \textit{explained by} phylogenetic networks.

In this work, we study the constraints imposed on orthology graphs when the underlying evolutionary
history follows a phylogenetic network instead of a tree. We show that any orthology graph can be
represented by a sufficiently complex level-k network. However, such networks lack biologically
meaningful constraints. In contrast, level-1 networks provide a simpler explanation, and we
establish characterizations for level-1 explainable orthology graphs, i.e., those derived from
level-1 evolutionary histories. To this end, we
employ modular decomposition, a classical technique for studying graph structures. Specifically, an arbitrary graph is level-1 explainable if and only
if each primitive subgraph is a near-cograph (a graph in which the removal of a single vertex
results in a cograph).
Additionally, we present a linear-time algorithm to recognize level-1 explainable orthology graphs
and to construct a level-1 network that explains them, if such a network exists.  Finally, we
demonstrate the close relationship of level-1 explainable orthology graphs to the substitution
operation, weakly chordal and perfect graphs, as well as graphs with twin-width at most 2.
}

\smallskip
\noindent

\section{Introduction}
In phylogenetics, the concept of homology describes genes that share a common ancestry, meaning they
diverged from a common ancestral gene in an ancestral species. Homology between two genes $x$ and $y$
can be further classified based on the type of evolutionary event that caused the divergence of
their most recent common ancestor.
If the divergence occurred due to speciation where the ancestral species split into two descendant
species, genes $x$ and $y$ are \emph{orthologs} \cite{Fitch:1970}. Homologous genes can also arise through evolutionary events
other than speciation, such as gene duplication or horizontal gene transfer. In these cases, the
genes are referred to as paralogs and xenologs, respectively \cite{Fitch:2000,Gray:1983}.

Orthologs are special in the sense that they represent gene copies found in different species. 
They often share similarities in sequence or structure and presumably perform similar functions in their respective
species. As a result, orthology data is a valuable resource in areas such as functional prediction \cite{Eisen:1998}, genetic diagnostics \cite{Forslund:2011} or even 
drug design \cite{Chandrasekaran:2016},  enabling researchers to transfer knowledge gained from genes in one species to their
orthologous counterparts in another.
Since the first definition of orthology in the early 1970s \cite{Fitch:1970}, inferring orthology information from genomic data has been an active field of research.
There are two main approaches to this task: tree-based reconciliation methods 
and graph-based inference methods, see \cite{Kuzniar2008,KWMK:11,Altenhoff2019} for an overview.

Tree-based methods typically require input gene and species phylogenies (trees or networks) and
reconcile these two phylogenies by computing a mapping from the gene phylogeny onto the species
phylogeny. While maximum likelihood \cite{Gorecki:2011} or Bayesian approaches \cite{Sjostrand:2014} exist, most methods employ
event-based maximum parsimony, see \cite{Menet:2022} for an overview. This involves assigning costs to certain evolutionary events
(e.g., duplication, speciation, and transfer) and finding a reconciliation that minimizes the total
cost of all events.
However, finding a cost-optimal reconciliation is computationally challenging \cite{Tofigh:2010},
making it impractical for genomwide data sets
containing thousands of genes. Additionally, tree-based methods are often sensitive to the input
phylogenies and the cost scheme used. Once a phylogenetic reconciliation is computed, the orthology
relationships can be directly inferred from the mapping.

Graph-based methods, in contrast, are used to infer an \emph{orthology graph}, where each vertex
corresponds to a gene, and an edge exists between two vertices if the corresponding genes are
estimated to be orthologs.
These methods aim to construct the orthology graph directly from
pairwise sequences comparisons, bypassing the need for explicit gene and species phylogenies.
Therefore, they are often more computationally efficient \cite{Altenhoff2019}.
Early graph-based methods inferred orthology by identifying best matches
or bi-directional best hits to compute clusters of orthologs \cite{Tatusov:1997,Dessimoz:2005}.
While effective as a starting point, these approaches do not
account for whether the underlying evolutionary history is tree-like or, more generally,
network-like.

This raises the question of whether such inferred orthology graphs are biologically
feasible, namely whether there exist evolutionary scenarios that support these estimations.
In the simplest setting, we may ask whether there exists a gene tree with internal nodes labeled as
1 (speciation) or 0 (non-speciation) such that two genes, $x$ and $y$, are linked by an edge in the
orthology graph if and only if their least common ancestor is labeled with 1. In this scenario, the
orthology graph is biologically feasible if and only if it is a so-called cograph, i.e., a graph
that does not contain an induced path on four vertices \cite{Hellmuth:2013, Hellmuth:2015}. In this case, the corresponding cotree represents a partially resolved version of the
underlying gene tree.
However, inferred orthology graphs often fail to satisfy the cograph property. One reason for this
may be noise in the data or limitations in the inference methods. Assuming a phylogenetic tree-like
model of evolution, we can leverage the structural property of true orthology graphs being cographs
to correct errors in the inferred orthology graph, for example, by editing it to the closest cograph
\cite{Hellmuth:2015,Dondi:2017}.
Another reason could be that the evolutionary history is not tree-like. Although phylogenetic trees
remain the most widely used evolutionary model, they are a simplification. It is well-acknowledged
that genes and species may evolve in a network-like manner due to events such as hybridization (at
the species level) or gene fusion (at the genetic level) \cite{Rivera:2004,Kummerfeld:2005}. Various classes of
phylogenetic networks have been studied, including galled trees, level-$k$ networks,
tree-child networks and many more \cite{Huson:2010,Kong:2022,Hellmuth2023}.

Thus, it is natural to ask what constraints a particular phylogenetic network model imposes on the
orthology graph. This question was addressed for trees \cite{Hellmuth:2013} and 
so-called galled trees in \cite{HS:22, LSH:24, HS:24,Huber2018} in which each vertex belongs
to at most one ``undirected cycle''. 
In this work, we extend the analysis to the broader class of level-$k$ networks, with a particular focus on the
prominent subclass of level-1 networks. We demonstrate that any orthology graph can be explained by
a level-$k$ network if $k$ is sufficiently large. Consequently, level-$k$ networks with arbitrary $k$ do
not impose specific constraints on orthology graphs. In contrast, level-1 networks do impose
constraints, revealing a close structural relationship to cographs.  

This paper is organized as follows. In Section~\ref{sec:prel}, we provide the basic definitions needed
throughout this work. We then continue in Section~\ref{sec:GeneralDAGs} to explain in more detail how orthology graphs
can be explained by DAGs or networks. We first review some of the results established to characterize those orthology graphs
that can be explained by labeled phylogenetic trees, which are cographs. Assuming that an orthology graph is not a 
cograph, it cannot be explained by a tree; however, there is always a set of vertices that, if removed from $G$, result in a cograph. 
We use this result to provide a novel construction for networks explaining orthology graphs. However, these networks
suffer from high complexity, in particular, they contain a high number of hybrid vertices, representing reticulate events. 
In Section~\ref{sec:lvl1ex}, we focus on orthology graphs that can be explained by networks with
relatively few hybrid vertices per ``block'', specifically level-1 networks. After a brief overview of
modular decomposition, a prominent and highly useful graph-theoretical tool in this context, we
first examine so-called primitive graphs. A primitive graph $G$ has only a ``trivial'' modular
decomposition and, in particular, cannot be written as a  join or a disjoint union of any
two graphs. In this sense, primitive graphs differ significantly from cographs, which are precisely
the graphs that can be constructed through a series of joins and disjoint unions.
Nevertheless, understanding primitive graphs that can be explained by level-1 networks provides the
foundation needed to characterize general level-1 explainable graphs. As
Theorem~\ref{thm:char-primitive-level1exp} demonstrates, a primitive graph can be explained by a
level-1 network if and only if it is a primitive near-cograph, meaning that there exists a vertex
whose removal results in a cograph. This result enables us to establish multiple characterizations
of general orthology graphs that can be explained by level-1 networks. More precisely, we prove that an
orthology graph $G$ is level-1 explainable if and only if
\begin{itemize}[noitemsep]
	\item[\ldots] every induced subgraph of $G$ is level-1 explainable (Proposition~\ref{prop:lev1ex-hereditary})
	\item[\ldots] for all non-trivial so-called prime modules in the modular decomposition of $G$, a well-defined quotient graph is a near-cograph (Theorem~\ref{thm:general-lev1ex})
	\item[\ldots] every primitive induced subgraph of $G$ is a near-cograph (Theorem~\ref{thm:genaral-lev1ex-v2})
	\item[\ldots] $G$ is explained by a \emph{phylogenetic} level-1 network (Theorem~\ref{thm:lev1ex<=>phylo-lev1-ex})
	\item[\ldots] $G$ can be constructed from a finite sequence of disjoint unions,  joins and vertex substitutions (Theorem~\ref{thm:char-LevIex-inflate} \& Theorem~\ref{thm:atomic-expression}).
\end{itemize}
Furthermore, we present a linear-time algorithm for recognizing level-1 explainable  orthology graphs
and to construct a level-1 network that explains them, if such a network exists (Theorem~\ref{thm:AlgGeneral}).
Finally, we show that the class of level-1 explainable graphs is a subclass of the class of
graphs with twin-width at most two, as well as the class of weakly chordal graphs, and
therefore, the class of perfect graphs. Since both perfect graphs and graphs of bounded
twin-width possess interesting algorithmic properties \cite{Grotschel:84,Bonnet:2024,Bonnet:2021A}, these results suggest that graphs explained by
level-1 networks may also be of interest beyond the scope of orthology graphs. We discuss this and
other potential directions for future research in the concluding Section~\ref{sec:outro}.

\section{Preliminaries}
\label{sec:prel}

\noindent
\paragraph{Sets}
All sets considered in this paper are assumed to be finite. We denote with $\binom{X}{2}$ the
two-element subsets of a non-empty set $X$ and with $2^X$ the powerset of $X$. A
subset $\mathfrak{S}\subseteq 2^X$ is also called a \emph{set system (on $X$)}.
For a given set system $\mathfrak{C}$  on $X$ and  a non-empty subset $W\subseteq X$, 
we define 
\[\mathfrak{C} \cap W \coloneqq  \{C\cap W \colon C\in \mathfrak{C} \text{ and } C \cap W\neq \emptyset\}. \]
In particular, for $x\in X$ we put $\mathfrak{C}-x \coloneqq \mathfrak{C}\cap (X\setminus\{x\})  =\{C\setminus \{x\} \colon C\in \mathfrak{C} \text{ and } C\neq \{x\}\}$.

A set system $\mathfrak{S}$ on $X$ is \emph{grounded} if $\emptyset\notin\mathfrak{S}$ and
$\{x\}\in\mathfrak{S}$ for each $x\in X$, and $\mathfrak{S}$ is a \emph{clustering system} if it is
grounded and $X\in \mathfrak{S}$. Two sets $X$ and $Y$ \emph{overlap} if $X\cap
Y\notin\{X,Y,\emptyset\}$. A clustering system that contains no two elements that overlap is
a \emph{hierarchy}.

Let $\mathfrak{C}$ be a clustering system on $X$. Then, 
$\mathfrak{C}$ on $X$ is \emph{closed} if, for all \emph{non-empty} $A\in
2^X$, it holds that $\bigcap_{\substack{C\in\mathfrak{C},\, A \subseteq C}} C =A \iff
A\in\mathfrak{C}$. A well known-characterization of closed  clustering system is as follows, cf.\ e.g.\ \cite[L.\ 16]{Hellmuth2023}
\begin{lemma}\label{lem:simple-closed}
  A clustering system $\mathfrak{C}$ is closed if and only if
  $A,B \in \mathfrak{C}$ and $A\cap B\ne\emptyset$ implies
  $A\cap B\in \mathfrak{C}$.
\end{lemma}

Furthermore, $\mathfrak{C}$ satisfies property (L) if
	  $C_1\cap C_2=C_1\cap C_3$ for all $C_1,C_2,C_3\in\mathfrak{C}$ where $C_1$
	  overlaps both $C_2$ and $C_3$.	  
$\mathfrak{C}$ is \emph{pre-binary} if, for all $x,y\in X$, there is a unique inclusion-minimal
element $C\in \mathfrak{C}$ such that $x,y \in C$. Moreover, $\mathfrak{C}$ is \emph{binary} if it
is pre-binary and, for all $C\in \mathfrak{C}$, there are $x,y\in X$ such that $C$ is the unique
inclusion-minimal element in $\mathfrak{C}$ with $x,y\in C$.

\noindent
\paragraph{Graphs, di-graphs, DAGs and networks}

An \emph{(undirected) graph} $G=(V,E)$ consists of a non-empty \emph{vertex set} $V(G)\coloneqq V$
and an \emph{edge set} $E(G)\coloneqq E \subseteq \binom{V}{2}$. 

\begin{remark}
We emphasize that every undirected graph $G$ can be interpreted as an orthology graph, that is, 
$V(G)$ represents a set of genes, and $\{x, y\} \in E(G)$ if and only if two distinct genes $x, y \in
V(G)$ are estimated to be orthologs. Therefore, instead of using the term ``orthology graph'',
we will simply use the term ``graph''.
\end{remark}

A \emph{subgraph} $H$ of $G$ is a graph $(W,F)$ that
satisfies $W\subseteq V$ and $F\subseteq E$. A subgraph $H$ of $G$ is \emph{induced} by $W\subseteq V$, 
denoted by $H =G[W]$, if $\{u,v\}\in E$ implies $\{u,v\}\in E(H)$ for all $u,v\in V(H)=W$. 
For a proper subset $W\subsetneq V(G)$, we denote with $G-W=G[V(G)\setminus W]$ the subgraph of $G$ induced by $V(G)\setminus W$
and use the simplified notation $G-v\coloneqq G-\{v\}$. A \emph{path} or \emph{$v_1v_k$-path} in $G$ is a 
is a sequence $v_1,\dots,v_k$ of pairwise distinct vertices satisfying $\{v_i,v_{i+1}\}\in E(G)$ for $i\in \{1,\dots,k-1\}$.
A path on four vertices is called a $P_4$. A graph is \emph{connected} if, for all vertices $u,v$ in $G$, there is a $uv$-path.

The \emph{complement $\overline G$} of a graph $G =(V,E)$ is the graph $\overline G \coloneqq (V, \{\{x,y\}\colon x,y \in V, x\neq y, \{x,y\}\notin E \} )$.
The \emph{join} of two vertex-disjoint graphs $H=(V,E)$ and $H'=(V',E')$ is defined by $H \join
H'\coloneqq (V\cup V',E\cup E' \cup \{\{x,y\}\mid x\in V, y\in V'\})$, whereas
their \emph{disjoint union} is given by $H \union H'\coloneqq (V \cup V',E \cup E')$.

A \emph{directed graph} $N=(V,E)$ consists of a non-empty vertex set $V(N)\coloneqq V$ and an edge set $E(N)\coloneqq E\subseteq \{(u,v)\mid u,v\in V \text{ and } u\neq v\}$.
A \emph{directed path} in a directed graph $N$ is a sequence $v_1,\dots,v_k$
of pairwise distinct vertices satisfying $(v_i,v_{i+1})\in E(N)$ for $i\in \{1,\dots,k-1\}$. We also call such a directed path a \emph{directed $v_1v_n$-path}. 
A \emph{directed acyclic graph (DAG)} $N$ is a directed graph that does not contain directed cycles,
that is, there is no directed $uv$-path such that $(v,u)\in E(N)$. 
A directed graph $N$ is \emph{connected} if its underlying undirected graph $N^U$ is connected, where
$N^U = (V(N), E(N^U))$ is obtained from $N$ by replacing each edge $(u,v)\in E(N)$
by the edge $\{u,v\}$.

Two undirected (resp., directed) graphs $G$ and $H$ are \emph{isomorphic}, in symbols $G\simeq H$, 
if there is a bijection $\varphi\colon V(G)\to V(H)$ such that for all $u,v\in V(G)$ it holds
that $\{u,v\}\in E(G)$ if and only if $\{\varphi(u),\varphi(v)\}\in E(H)$ 
(resp., $(u,v)\in E(G)$ if and only if $(\varphi(u),\varphi(v))\in E(H)$).

We can associate a partial order $\preceq_N$ on the vertex set $V(N)$ with a DAG $N$, defined by
$v\preceq_N w$ if and only if there is a directed $wv$-path in $N$. In this case, we say that $w$ is an
\emph{ancestor} of $v$ and $v$ is a \emph{descendant} of $w$. If $v\preceq_N w$ and $v\neq w$, we
write $v\prec_N w$. If $(u,v) \in E(N)$, then $u$ is a \emph{parent} of $v$ and $v$ a \emph{child}
of $u$. The sets $\parent_N(v)$ and $\child_N(v)$ comprise all parents and children of $v$,
respectively. Two vertices $u,v\in V(N)$ are \emph{$\preceq_N$-comparable} if $u\preceq_N v$ or $v\preceq_N u$ and, otherwise, $u$ and $v$ are \emph{$\preceq_{N}$-incomparable}.

Let $N=(V,E)$ be a DAG.
A vertex with $v\in V$ is a \emph{leaf} if $\outdeg_N(v)=0$, 
a \emph{root} if $\indeg_N(v)=0$,
a \emph{hybrid vertex} if $\indeg_N(v)>1$, and \emph{tree vertex} if
$\indeg_N(v)\le 1$. 
Equivalently, a leaf of $N$ is a $\preceq_N$-minimal vertex, while a root of $N$ is a $\preceq_N$-maximal vertex.
The set of leaves is denoted by $L(N)$. 
Moreover, the vertices that are not leaves are called \emph{inner vertices}, and
comprised in the set
$V^0(N)\coloneqq V(N)\setminus L(N)$.
A DAG $N$ \emph{on $X$} is a DAG with leaf-set $L(N)=X$.
Note that leaves and inner vertices can be hybrid vertices or tree vertices.

Following \cite{Hellmuth2023}, we define (phylogenetic) networks here
as a slightly more general class of DAGs than what is customarily considered in most of the literature on the
topic. 
\begin{definition}
  \label{def:N}
  A \emph{network} is a DAG $N=(V,E)$ such that
  \begin{itemize}
  \item[(N1)] There is a unique root of $N$, denoted by $\rho_N$.
  \end{itemize}
  A network $N$ is \emph{phylogenetic} if
  \begin{itemize}
  \item[(N2)] There is no vertex $v\in V$ such that  $\outdeg_N(v)=1$ and $\indeg_N(v)\le 1$.
  \end{itemize}
\end{definition}

For every vertex $v\in V(N)$ in a DAG $N$, the set of its descendant leaves
$\CC_N(v)\coloneqq\{ x\in L(N)\mid x \preceq_N v\}$
is a \emph{cluster} of $N$. We write $\mathfrak{C}_N\coloneqq\{\CC_N(v)\mid v\in V(N)\}$ for the set
of all clusters in $N$. Note that $\mathfrak{C}_N$ is a grounded set system on
$L(N)$ for every DAG $N$ \cite{LH:24a}. Moreover, $\mathfrak{C}_N$ is a clustering system for every network $N$; 
cf.\ \cite[Lemma 14]{Hellmuth2023}.

For a given a DAG $N$ and a non-empty subset $A\subseteq L(N)$, a vertex $v\in V(N)$ is a \emph{common
ancestor of $A$} if $v$ is ancestor of every vertex in $A$. Moreover, $v$ is a \emph{least common
ancestor} (LCA) of $A$ if $v$ is a $\preceq_N$-minimal common ancestor of
$A$. The set $\LCA_N(A)$ comprises all LCAs of $A$ in $N$. 
We will, in particular, be interested in situations where $|\LCA_N(A)|=1$ holds for certain subsets
$A\subseteq X$. For simplicity, we will write $\lca_N(A)=v$ in case that $\LCA_N(A)=\{v\}$  and say that
\emph{$\lca_N(A)$ is well-defined}; otherwise, we leave $\lca_N(A)$ \emph{undefined}. 
Moreover, we write $\lca_N(x,y)$ instead of $\lca_N(\{x,y\})$.

\begin{lemma}[{\cite[{L.~1 \& Obs.~2}]{Shanavas2024}}]\label{lem:lca-cluster}
Let $N$ be a network on $X$. If $u,v\in V(N)$ satisfy $u\preceq_N v$, then 
$\CC_N(u)\subseteq \CC_N(v)$. 
Moreover, if $\lca_N(A)$ is well-defined for some non-empty subset $A\subseteq X$, 
then $\lca_N(A) \preceq_N v$ for all $v$ with $A\subseteq \CC_N (v)$ and 
$\CC_N ( \lca_N (A))$ is the unique inclusion-minimal cluster in
$\mathfrak{C}_N$ containing $A$.
\end{lemma}

The previous lemma highlights the connection between clusters in a network and the least common ancestor map.
We now define specific properties of DAGs that rely on the LCAs and that play a central role in this
contribution.

\begin{definition}
Let $N$ be a DAG on $X$. Then, $N$ has the \emph{$k$-lca property} if $\lca_N(A)$ is well-defined for all nonempty subsets $A\subseteq X$ of size $|A|\leq k$. 
A vertex $v\in V(N)$ is a
\emph{2-lca vertex} if $v = \lca_N(A)$ for some subset $A\subseteq X$ of size $|A|\leq 2$. Moreover,
$N$ is \emph{2-lca-relevant} if each vertex $v\in V(N)$ is a 2-lca vertex.

If $N$ is a network with  $|X|$-lca property, then it is simply called \emph{lca-network}.
\end{definition}

\noindent
\paragraph{Blocks and level-$k$ networks}

A directed graph is \emph{biconnected} if it contains no
vertex whose removal disconnects the graph. 
Note that biconnected directed graphs are necessarily connected.
A \emph{block} in
a directed graph is an (inclusion) maximal biconnected subgraph. 
A block $B$ is called \emph{non-trivial} if it is not a
single vertex or a single edge.

Note that a non-trivial block $B$ in a DAG $N$ may have several $\preceq_N$-maximal
elements. This situation changes in networks. We summarize here some useful results
for later reference.

\begin{lemma}[{\cite[L.~8, 9 \& 18, Obs.~1]{Hellmuth2023}}]\label{lem:unique-maxB} 
Every block $B$ in a network $N$ has a unique $\preceq_N$-maximal vertex $\max_B$. 
In particular, for every $w \in  V(B)$, there is a directed path from $\max_B$ to $w$ 
and every such path is completely contained in $B$. 

If $u$ and $v$ are distinct vertices of a block $B$ and a block $B'$, 
then $B=B'$. If $v$ is contained in block $B$ and $B'$ but $v\notin \{\max_B, \max_{B'}\}$, 
then $B=B'$. 

If $u$ and $v$ are $\preceq_N$-incomparable vertices in $N$ such that 
$\CC_N(v)\cap \CC_N(u)\neq \emptyset$, then $u$ and $v$ are contained in a common
non-trivial block $B$ in $N$. In this case, there is a hybrid vertex $h$ in 
$N$ that satisfies $h\prec_N u$ and  $h\prec_N v$
\end{lemma}

\begin{definition}
  \label{def:level-k}
  A network $N$ is \emph{level-$k$} if each block $B$ of $N$ contains at
  most $k$ hybrid vertices distinct from $\max_B$.
\end{definition}

A level-0 network is a \emph{tree}. 
In what follows, we are mainly interested in trees and level-1 networks.
Lemma 50 together with Corollary 21 in \cite{Hellmuth2023} imply

\begin{lemma}\label{lem:lvl1=>C-is-binary}
For every level-1 network $N$ the clustering system $\mathfrak{C}_N$ is binary.
\end{lemma}

\begin{lemma}[{\cite[L.~49, Thm.~8]{Hellmuth2023}}]\label{lem:lvl1=>lcaN}
If  $N$ is a level-1 network, then $\mathfrak{C}_N$ is closed and
$N$ is an lca-network.
\end{lemma}

Lemma~\ref{lem:lvl1=>lcaN} implies, in particular, that every level-1 network
has the 2-lca property.

\begin{lemma}\label{lem:lvl1-intersections}%
Let $N$ be a phylogenetic level-1 network and $u,v\in V(N)$. 
Then, one of the following statements is true: 
\begin{enumerate}[label=(\roman*)]
\item $\CC_N(u)\subseteq \CC_N(v)$ or $\CC_N(v)\subseteq \CC_N(u)$ or 
\item $\CC_N(u)\cap \CC_N(v) = \emptyset$ or $\CC_N(u)\cap \CC_N(v) = \CC_N(h)$ for some hybrid vertex $h \in  V(N)$ 
\end{enumerate}
\end{lemma}
\begin{proof}
	If $u$ and $v$ are $\preceq_N$-comparable, then Lemma~\ref{lem:lca-cluster}
	implies that Condition (i) holds. 
	Suppose now that $u$ and $v$ are $\preceq_N$-incomparable. 
	By \cite[Prop.\ 9]{Hellmuth2023}, every phylogenetic level-1
	network is a so-called tree-child network.
	This and \cite[L.~34]{Hellmuth2023} implies that 
	either  $\CC_N(u)\cap \CC_N(v) = \CC_N(h)$ for some hybrid vertex $h \in  V(N)$
	or  $\CC_N(u)\cap \CC_N(v) \notin \mathfrak{C}_N$. 
	By Lemma~\ref{lem:lvl1=>lcaN}, $\mathfrak{C}_N$ is closed for 
	every  level-1 network $N$ and Lemma~\ref{lem:simple-closed}
	implies that, in case that $\CC_N(u)\cap \CC_N(v) \notin \mathfrak{C}_N$, 
	$\CC_N(u)\cap \CC_N(v) = \emptyset$ must hold. Hence, 
	Condition (ii) holds. 	
\end{proof}

\noindent	
\paragraph{Hasse diagram and regular DAGs}

If $\mathfrak{C}$ is a set system on $X$, then the \emph{Hasse diagram} $\hasse(\mathfrak{C})$ (also
known as the \emph{cover digraph} of $\mathfrak{C}$ \cite{sem-ste-03a}) is the DAG with vertex set $\mathfrak{C}$ and directed
edges $(A,B)$ for each pair of vertices $A,B\in\mathfrak{C}$ such that (i) $B\subsetneq A$ and (ii)
there is no $C\in\mathfrak{C}$ such that $B\subsetneq C\subsetneq A$. A DAG $N=(V,E)$ is
\emph{regular} if the map $\varphi\colon V\to V(\hasse(\mathfrak{C}_N))$ defined by $v\mapsto \CC_N(v)$
is an isomorphism of $N$ and $\hasse(\mathfrak{C}_N)$ \cite{Baroni:2005}. 
We summarize now some the structural properties of regular DAGs and networks.

\begin{lemma}\label{lem:regular-props}
	If $N$ is a regular DAG, then $N$ is phylogenetic and satisfies $\CC_N(v)\neq\CC_N(u)$ for all distinct $u,v\in V(N)$.
\end{lemma}
\begin{proof}
	Let $N$ be a regular DAG. Theorem~4.6 of \cite{LH:24a} states that $N$ is phylogenetic (c.f. \cite[Thm.~2]{Hellmuth2023} for networks). Since $\varphi\colon V(N)\to
	\mathfrak{C}_N$ is an isomorphism between $N$ and $\hasse(\mathfrak{C}_N)$,
	the map $\varphi$ is, in particular, injective. Thus, 
	$u\neq v$ implies $\CC_N(u)=\varphi(u)\neq\varphi(v)=\CC_N(v)$ for all $u,v\in V(N)$.
\end{proof}

One non-appealing property of the Hasse diagram $\hasse(\mathfrak{C})$ of a grounded set system
$\mathfrak{C}$ on $X$ is the fact that the leaves of $\hasse(\mathfrak{C})$ consist of the inclusion-minimal
elements of $\mathfrak{C}$, that is, of the singletons $\{x\}$ for each $x\in X$. Consequently, we
have $\mathfrak{C}\neq\mathfrak{C}_{\hasse(\mathfrak{C})}$. For practical reasons, we thus write
$H\doteq\hasse(\mathfrak{C})$ for the directed graph that is obtained from $\hasse(\mathfrak{C})$ by
relabeling all vertices $\{x\}$ in $\hasse(\mathfrak{C})$ by $x$. Whenever $\mathfrak{C}$ is a
grounded set system, $H\doteq\hasse(\mathfrak{C})$ thus satisfies $\mathfrak{C}_H = \mathfrak{C}$. 

For later reference, we provide two interesting results showing the connection of clustering
systems of trees and level-1 networks with respect to their their underlying Hasse diagrams.

\begin{theorem}[{\cite[Sec.~3.5]{sem-ste-03a} and \cite[Cor.~9]{Hellmuth2023}}]\label{thm:CharTreeCluster}
	Let $\mathfrak{C}$ be a clustering system  on $X$. Then, there is a 
	phylogenetic tree $T$ on $X$ such that $\mathfrak{C}_T = \mathfrak{C}$
	if and only if $\mathfrak{C}$ is a hierarchy.
	In this case, $H\doteq \Hasse(\mathfrak{C})$ is a phylogenetic tree on $X$. 
\end{theorem}

\begin{theorem}[{\cite[Prop.~18, Cor.~36]{Hellmuth2023}}]\label{thm:CharLvl1Cluster}
	Let $\mathfrak{C}$ be a clustering system  on $X$. Then, there is a 
	phylogenetic level-1 network $N$ on $X$ such that $\mathfrak{C}_N = \mathfrak{C}$
	if and only if $\mathfrak{C}$ is a closed and satisfies (L). 
	In this case, 	$H\doteq \Hasse(\mathfrak{C})$ is a phylogenetic level-1 network on $X$. 
\end{theorem}

\section{Graphs and orthologs explained by DAGs, networks and trees}
\label{sec:GeneralDAGs}

We will consider \emph{labeled} DAGs $(N,t)$, that is, DAGs $N=(V,E)$ equipped with a
(vertex-)\emph{labeling} $t\colon V^0(N)\to\Upsilon$ where $\Upsilon$ denotes some set of labels.
In case that $\Upsilon =\{0,1\}$ we call $(N,t)$ also \emph{$0/1$-labeled} and $t$ a \emph{$0/1$-labeling}.
If $N$ is a DAG on $X$ that has the 2-lca property, 
then we can derive from  any $0/1$-labeled version $(N,t)$ the 
graph 
\begin{center}
$\mathscr{G}(N,t) =(X,E)$, where $\{x,y\}\in E$ if and only if $t(\lca_N(x,y))=1$ and $x,y\in X$. 
\end{center}

\begin{definition}\label{def:strudi-explain}
Let $(N,t)$ be $0/1$-labeled DAG on $X$ and assume that $N$ has the 2-lca property.
A graph  $G = (X,E)$ is \emph{explained by} $(N,t)$ if $G = \mathscr{G}(N,t)$. 
For the trivial case $|X|=1$, we say that $G=(X,\emptyset)$ is explained by 
$(N,t)$ where $N=(X,\emptyset)$ is the single vertex graph and $t$ is arbitrary with empty domain. 
\end{definition}

Examples illustrating Definition~\ref{def:strudi-explain} are provided in Figure~\ref{fig:someN} and \ref{fig:level2counterex}.
Clearly, if $G = \mathscr{G}(N,t)$, then its complement $\overline G$
is explained by the $0/1$-labeled network $(N,\overline t)$ for which $\overline t(v) = 1$ if and only if $t(v)=0$ for all $v\in V^0(G)$. 
	For later reference, we summarize this into 
	\begin{observation}\label{obs:complement}
		$G = \mathscr{G}(N,t)$ if and only if $\overline G = \mathscr{G}(N,\overline t)$. 
	\end{observation}

\subsection{Existing constructions}

An important role in this contribution is played by graphs that can be explained by a $0/1$-labeled
tree, namely, cographs. The class of \emph{cographs} is the smallest class of graphs that contains
the single-vertex graph and is closed under the join- and disjoint union operators (also known as
series and parallel composition) \cite{Corneil:85}. Cographs can be characterized as follows.

\begin{theorem}[{\cite{Corneil:81,Sumner74,Seinsche:74}}]\label{thm:CharCograph}
Given a graph $G$, the following statements are equivalent:
\begin{enumerate}[label=(\roman*)]
	\item $G$ is a cograph.
	\item $\overline G$ is a cograph.
	\item $G$ does not contain an induced path $P_4$ on four vertices.
	\item Every induced subgraph of $G$ is a cograph.
	\item $G = \mathscr{G}(T,t)$ for some $0/1$-labeled tree $(T,t)$. 
	
	In particular, there is a unique (up to isomorphism) $0/1$-labeled phylogenetic tree
	in which no two adjacent vertices of the tree have the same label, called \emph{cotree}, that explains $G$.
\end{enumerate}
\end{theorem}

\begin{figure}[tbp]
  \centering
  \includegraphics[width=0.7\textwidth]{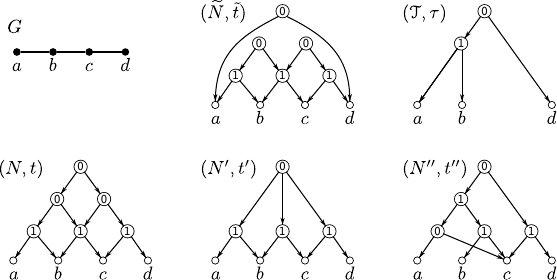}
  \caption{	Shown are four $0/1$-labeled DAGs $(N,t)$, $(N',t')$,  $(N'',t'')$ and $(\widetilde N, \tilde t)$ that all explain the graph $G$.
  				Here, $N$, $N'$ and $N''$ are networks while $\widetilde N$ is not.
  				Since $G\simeq P_4$, Theorem~\ref{thm:CharCograph} implies
  				that $G$ cannot be explained by a $0/1$-labeled tree. 
  				The network $(N,t)$ is a ``half-grid'' (cf.\ \cite{BSH:22}) and is level-$3$, whereas
  				$(N',t')$ is a regular level-2 network with $N'\doteq\Hasse(\mathfrak{C}_G)$ where $\mathfrak{C}_G$ is chosen according to Equation~\eqref{eq:C_G}.
  				The network $(N'',t'')$ is a level-1 network and coincide with  $(\MDT_{\NWA c},\tau_{\NWA c})$
  				that is obtained from the cotree $(\MDT,\tau)$ (shown in the upper right corner)	of the cograph $G-c$
  				 (cf.\ Definition~\ref{def:graft} and Theorem~\ref{thm:N<-v_explains-G}).
  				}
  \label{fig:someN}
\end{figure}

By Theorem~\ref{thm:CharCograph}, any graph that contains an induced $P_4$ cannot be explained by a
$0/1$-labeled tree. This naturally raises the question of which types of graphs can be explained by
which types of $0/1$-labeled networks.
In \cite{BSH:22}, it was shown that every graph can be explained by a so-called $0/1$-labeled
half-grid (see the network $(N, t)$ in Figure~\ref{fig:someN} for an illustrative example). However,
such half-grids are rather complex, as their level is $k \in \Omega(|X|^2)$.

A simpler construction, closely related to the one provided in the proof of \cite[Prop.~3.3]{LSH:24}, is as follows. 
For a given graph $G = (X, E)$, define the set system 
\begin{equation}\label{eq:C_G}
\mathfrak{C}_G \coloneqq \{\{x, y\} \colon x, y \in X \text{ and } (x = y \text{ or } \{x, y\} \in E)\} \cup \{X\}.
\end{equation}
By construction, $\mathfrak{C}_G$ is a clustering system. Now, consider the network $N \doteq \Hasse(\mathfrak{C}_G)$ and a 2-element subset $\{x, y\}$ of $X$.  
If $\{x, y\} \in E$, then $\lca_N(x, y) = v$, where $v$ is the unique vertex such that $\CC_N(v) = \{x, y\}$.  
If $\{x, y\} \notin E$, then $\lca_N(x, y) = \rho_N$. In particular, for every subset $Y \subseteq X$, 
$\lca_N(Y)$ is well-defined, i.e., $N$ is an lca-network. Since $N$ is regular, Lemma~\ref{lem:regular-props} implies that $N$ is phylogenetic.
Moreover, we can equip $N$ with the $0/1$-labeling $t$ defined by $t(v) = 1$ for all $v \in V^0(N) \setminus \{\rho_N\}$ 
and $t(\rho_N) = 0$ to obtain the network $(N, t)$ that explains $G$ (see the network $(N',t')$ in Figure~\ref{fig:someN} for an
illustrative example). Furthermore, any hybrid in $N$ must be a leaf, 
implying that $N$ is level-$k$ for $k \in O(|X|)$.
The latter, in particular, implies 
\begin{proposition}\label{prop:G-hasExpl}
	For every graph  $G = (X,E)$ there is a $0/1$-labeled phylogenetic regular level-$k$ lca-network $(N,t)$ with
	$O(|X|+|E|)\subseteq O(|X|^2)$ vertices and $k\in O(|X|)$ that explains $G$. 
\end{proposition}

Proposition~\ref{prop:G-hasExpl} implies that every sparse graph $G = (X, E)$, 
i.e., a graph with $|E| \in O(|X|)$, can be explained by a network $(N, t)$ having 
$O(|X|)$ vertices. Moreover, for some dense graphs $G = (X, E)$, i.e., graphs with $|E| \in \Omega(|X|^2)$,  
their complement $\overline{G} = (X, \overline{E})$ is sparse. In this case, 
Proposition~\ref{prop:G-hasExpl}, together with Observation~\ref{obs:complement}, implies that 
these types of graphs can also be explained by networks $(N, t)$ having $O(|X|)$ vertices. This property, 
however, is not guaranteed if both $G$ and $\overline{G}$ are dense.

Proposition~\ref{prop:G-hasExpl} implies, in particular, that the property of a graph being explainable by a
$0/1$-labeled lca-network is hereditary. As demonstrated later in Proposition~\ref{prop:lev1ex-hereditary},
this hereditary property also applies to graphs explained by level-1 networks. The construction of a
$0/1$-labeled lca-network that explains induced subgraphs of a graph $G$,
based on specific clusters in the network explaining $G$, is presented in the following result.

\begin{proposition}[{\cite[Prop.~3.4]{LSH:24}}]\label{prop:lcaN-explain}
	Let  $(N,t)$ be a $0/1$-labeled lca-network that explains $G$ and $W\subseteq V(G)$ be a non-empty subset. 
Put  $\mathfrak{C}'\coloneqq \mathfrak{C}_N\cap W$.
Then, $\mathfrak{C}'$ is a clustering system on $W$ and 
	$N'\doteq \Hasse(\mathfrak{C}')$ is an lca-network.  Moreover,  $(N',t')$ explains $G[W]$,
	where $t'$ is the labeling defined by putting $t'(v)\coloneqq t(\lca_N(\CC_N(v)))$ for all inner vertices $v$ of $G$.
\end{proposition}

\subsection{Novel constructions}

In what follows, we provide further methods to construct 
level-$k$ networks, with $k < |X|$, that can explain a given graph $G = (X, E)$. 
To this end, we first consider subsets of clustering systems, denoted by $\mathfrak{C}(\{1,2\})$,  
associated with networks that explain a given graph, and their Hasse diagram $\Hasse(\mathfrak{C}(\{1,2\}))$. 
More precisely, in \cite{LH:24a,LH:24b} DAGs that are 2-lca-relevant have been studied. These DAGs have strong 
connections to regular DAGs and can, in particular, be computed using a certain $\ominus$-operator 
that acts on the vertices and edges of a given DAG. For brevity, we omit the details of the 
$\ominus$-operator and instead summarize the most important results from \cite{LH:24a,LH:24b} in the following theorem
which, in addition, includes new results on networks explaining graphs.

\begin{theorem}\label{thm:regular-version}
	Let $G=(X,E)$ be a graph explained by a $0/1$-labeled network $(N,t)$ and define
	\[\mathfrak{C}(\{1,2\})\coloneqq \left\{C\in\mathfrak{C}_N\,\colon\, 
	 C \text{ is the unique inclusion-minimal cluster in } 
	\mathfrak{C}_N \text{ containing }
	A \text{ for some }  A\subseteq X,\, |A|\in\{1,2\} \right\}.\]
	Then, $\Hasse(\mathfrak{C}(\{1,2\}))$ is a regular DAG that is 2-lca-relevant.
	In particular, there is a DAG $H \simeq \Hasse(\mathfrak{C}(\{1,2\}))$ on $X$
	such that $V(H)\subseteq V(N)$ and  $\lca_N(x,y)=\lca_{H}(x,y)$ for all $x,y\in X$.
	In addition, $H$  can be equipped with a labeling $t'$ such that $(H,t')$ explains $G$.
	Moreover, if $N$ is level-1, then $H$ is a phylogenetic level-1 network.
\end{theorem}
\begin{proof}
	Let $G=(X,E)$ be a graph explained by a $0/1$-labeled network $(N,t)$ and $\mathfrak{C}(\{1,2\})$ be defined as
	in the statement of the theorem. Note that $N$ must have the $2$-$\lca$ property. Thus, we can 
	apply \cite[Thm.~3.12]{LH:24b} which states that $H \simeq \Hasse(\mathfrak{C}(\{1,2\}))$
	is a regular and 2-lca relevant DAG
	on $X$ that satisfies  $V(H)\subseteq V(N)$ and $\lca_N(x,y)=\lca_{H}(x,y)$ for all $x,y\in X$.
	Since $V(H)\subseteq V(N)$, we can define $t'\coloneqq V^0(H) \to \{0,1\}$ 
	by putting $t'(v) \coloneqq t(v)$ for all $v\in V^0(H)$.
	Since $\lca_N(x,y)=\lca_{H}(x,y)$ for all $x,y\in X$, we can conclude that 
	$t(\lca_{N}(x,y))=t'(\lca_H(x,y))$ for all $x,y\in X$. Hence, 
	$\mathscr{G}(H,t')=G$ and it follows that $(H,t')$ explains $G$.

	Finally assume that $N$ is a level-1 network.  
	By Lemma~\ref{lem:lvl1=>C-is-binary}, $\mathfrak{C}_N$ is binary. Hence,  
	$\mathfrak{C}(\{1,2\})=\mathfrak{C}_{N}$. Theorem~\ref{thm:CharLvl1Cluster}
	implies that $\mathfrak{C}(\{1,2\})$ is closed and satisfied (L)
	and that $H$ is a phylogenetic level-1 network.
 \end{proof}

\begin{figure}
  \centering
  \includegraphics[width=0.65\textwidth]{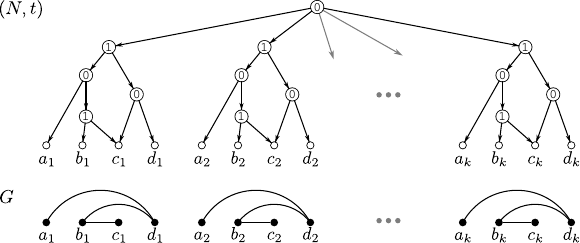}
  \caption{A $0/1$-labeled level-1 network $(N,t)$ where $G=\mathscr{G}(N,t)$ consists of $k>1$ vertex disjoint induced $P_4$s.  
  		If $G-Y$ is a cograph, then $Y\subseteq V(G)$ with $|Y|\geq k>1$ must hold.}
  \label{fig:level2counterex}
\end{figure}

Theorem~\ref{thm:regular-version} and Proposition~\ref{prop:lcaN-explain} are closely related, as
both provide methods for constructing labeled DAGs and networks that explain a given graph $G$ or a subgraph
of $G$. Theorem~\ref{thm:regular-version} shows that any graph explained by a $0/1$-labeled network
can also be explained by a regular 2-lca-relevant DAG, with additional guarantees when the network
is level-1. In contrast, Proposition~\ref{prop:lcaN-explain} specifically requires the input network
to be an lca-network and focuses on constructing a new lca-network that explains an induced subgraph
$G[W]$. Thus, while both results address network constructions, Proposition~\ref{prop:lcaN-explain}
imposes a stronger assumption on the input network.

In the following, we present a simple method for extending a DAG $N$ by introducing a new vertex $z$
and applying a series of operations, resulting in the DAG $N_{\NWA z}$. This procedure is essential
for extending a $0/1$-labeled network to explain a larger graph while preserving its key structural
properties. Definition~\ref{def:graft} provides the formal details for constructing
$N_{\NWA z}$. Next, Lemma~\ref{lem:graft-properties} establishes important properties of
$N_{\NWA z}$, including the preservation of network characteristics (e.g., phylogenetic or the 2-lca
property) and showing that the level of $N_{\NWA z}$ increases by at most one compared to the
number of hybrids of $N$. Finally, Theorem~\ref{thm:N<-v_explains-G} demonstrates that for a graph
$G - z$ explained by $(N,t)$, the network $N_{\NWA z}$ can be equipped with a $0/1$-labeling
$t_{\NWA z}$, resulting in a labeled network $(N_{\NWA z}, t_{\NWA z})$ that explains the entire
graph $G$.

\begin{definition}\label{def:graft}
Let $N$ be a connected DAG on $X$ and $v\in V(N)$. The \emph{expanding operation $\expand(v)$} is
defined as follows: create a new vertex $w_v$, replace edges $(u, v)$ by $(u, w_v)$ for all $u \in
\parent_N (v)$, and add the edge $(w_v, v)$. 

\smallskip
\noindent
Let $z\notin V(N)$ be a vertex that is not part of $N$. If $|X|>1$,
then the directed graph $N_{\NWA z}$ is obtained from $N$ by 
	  \begin{enumerate}
	  	\item applying $\expand(x)$ on every $x\in X$ with $\indeg_N(x)> 1$ and, afterwards, \label{graft:leafextend}
	  	\item replacing every edge $(u,x)$ by the two edges $(u,u_x)$ and $(u_x,x)$ for all $x\in X$ and, afterwards, \label{graft:subidiv}
	  	\item adding the new vertex $z$ together with edge $(u_x,z)$ for all $x\in X$. \label{graft:z}
	 \end{enumerate}
\noindent
If $|X|=1$, then $X =V(N)$, and we define  $N_{\NWA z}$ as the network with a single root $u_x$
that has precisely the two children $z$ and $x\in X$ and no other vertices or edges.
\end{definition}

\begin{figure}
  \centering
  \includegraphics[width=0.9\textwidth]{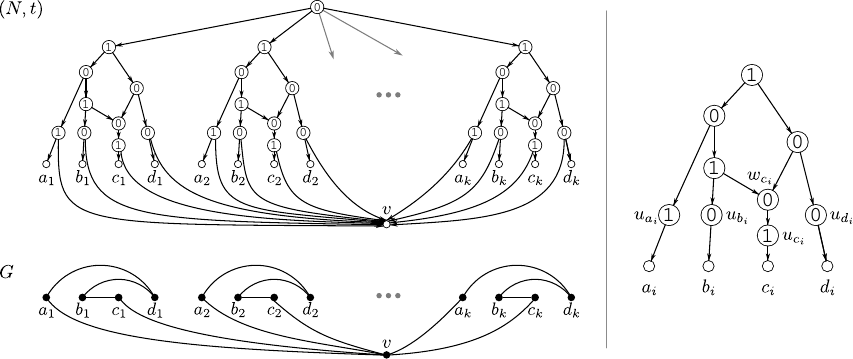}
  \caption{
 	 Example for Definition~\ref{def:graft}, constructing the shown $0/1$-labeled network $(N_{\NWA v},t_{\NWA v})$ from the level-1 network $(N,t)$ as shown in Figure~\ref{fig:level2counterex}. 
 	 The graph $G-v = \mathscr{G}(N,t)$ is explained by the level-1 network $(N,t)$, however, 
 	 the network $(N_{\NWA v},t_{\NWA v})$ is level-$(k+1)$, $k>1$. 
 	 Note that $G$ contains $k$ induced cycles $C_5$ on five vertices and is, by Lemma~\ref{lem:odd-hole-free},
 	 not \levIex.}
 	 \label{fig:graft}
\end{figure}

In the following, we always assume that for the constructed directed graph $N_{\NWA z}$ it holds that  $z\notin V(N)$.
Moreover, observe that $V(N) \subseteq V(N_{\NWA z})$ and $V(N_{\NWA z})\setminus V(N) = \{z\} \cup \{u_x\colon x\in X\} \cup \{w_x \colon x\in X \text{ is a hybrid in } N  \}$.
Definition~\ref{def:graft} is illustrated in Figure~\ref{fig:graft}. 

\begin{lemma}\label{lem:graft-properties}
	Suppose that $N$ is a DAG on $X$ with the 2-lca property.
	Then,  $N_{\NWA z}$ is a DAG on $X\cup\{z\}$ with the 2-lca property that satisfies, for all 
	$x,y\in X\cup\{z\}$, 
	\begin{equation}\label{eq:lca-NWA}
	\lca_{N_{\NWA z}}(x,y)=
		\begin{cases}
		 \lca_N(x,y)&\text{if } z\notin\{x,y\}\\
		 u_x & \text{if } z=y\\
		 u_y & \text{otherwise, i.e., } z=x. 
		\end{cases}
	\end{equation}	
Moreover, if $N$ is a network, then $N_{\NWA z}$ is a network and, 
if  $N$ is phylogenetic, then  $N_{\NWA z}$ is phylogenetic. 
In particular, if $N$ is an $\lca$-network, then $N_{\NWA z}$ is an $\lca$-network.
Furthermore, if $N$ is a network that has at most $k-1$ hybrid vertices, then $N_{\NWA z}$ has at most $k$ hybrids and is level-$k$.
\end{lemma}
\begin{proof} 
	Let $N$ be a DAG on $X$ with the 2-lca property. Clearly, $N$ must be connected since if not, then at least two leaves have no common ancestor at all. 
	It is straightforward to verify that the statements hold
	in case that $|X|=1$. Hence, suppose in the following that $|X|>1$. This and connectedness of $N$ implies that $|E(N)|>1$. To simplify writing, put $N'\coloneqq N_{\NWA z}$
	where $N_{\NWA z}$ is defined according to Definition~\ref{def:graft}. One easily
	observes that $N'$ is a DAG with leaf-set $X\cup\{z\}$.
	Note that by construction, $V(N)\subseteq V(N')$ and $w\preceq_N w'$ if and only if
	$w\preceq_{N'}w'$, for all $w,w'\in V(N)$. In particular, each new vertex
	$u\in V(N')\setminus V(N)$ satisfies $u=w_x$ or $u=u_x$
	for some $x\in X$ or $u=z$ according to Step~\eqref{graft:leafextend}, \eqref{graft:subidiv} and \eqref{graft:z}
	in Definition~\ref{def:graft}, respectively. Thus, 
	$V(N')\setminus V(N) = \{z\} \cup \{u_x\colon x\in X\}\cup\{w_x\colon x\in X\text{ and } \indeg_N(x)>1\}$.
	By construction, $\CC_{N'}(u)=\{z,x\}$ for all $u\in V(N')\setminus (V(N) \cup \{z\})$  and $\CC_{N'}(z)=\{z\}$.
	Therefore, no vertex in $V(N')\setminus V(N)$ can be a common ancestor of two distinct elements in $X$. 
  
	Now let $x,y\in X$. Since $N$ has the 2-lca property, $\lca_N(x,y)$ is well-defined. As previously
	argued, there is no common ancestor of $x$ and $y$ contained in $V(N')\setminus V(N)$, and since
	the partial order of $N$ agrees with that of $N'$ on common ancestors of $x$ and $y$ in $N$, we can
	conclude that $\lca_N(x,y)=\lca_{N'}(x,y)$ for all $x,y\in X$. Next, consider $z$ and some $x\in X$. 
	There is, by construction of $N'$, a unique parent $u_x$ of $x$ in $N'$. Moreover, $(u_x, z)$
	is  an edge of $N'$, hence $u_x$ is a common ancestor of $z$ and $x$ in $N'$. Since
	$u_x$ is the unique parent of $x$ in $N'$, every other common ancestor $u$ of $z$ and $x$ in $N'$
	satisfy $u_x\prec_{N'} u$. Thus, $\lca_{N'}(x,z)$ is well-defined and satisfies $\lca_{N'}(x,z)=u_x$
	for all $x\in X$. In summary, $N'$ has the 2-lca property.

	Clearly, if $N$ is a network, then the root $\rho$ of $N$ remains the unique root in $N'$, i.e.,
    $N'$ is a network. Now, assume that $N$ is phylogenetic. By construction of
    $N'$, the only vertices that are in $V(N)\cap V(N')$ and that differ in their in- or out-degrees are
    those hybrids $x$ in $N$ that are leaves. Such vertices still have out-degree zero in $N'$. Moreover,
    $z$ has $\indeg_{N'}(z)>1$ since $|X|>1$. To recall, each new vertex $u\in V(N')\setminus(V(N)\cup\{z\})$ satisfies 
	$u=u_x$ or $u=w_x$ for some $x\in X$. If $u=u_x$, then $u$ has exactly two children $z$ and $x$ and thus, $\outdeg_{N'}(u)>1$. 
	If $u=w_x$ then $\indeg_{N'}(w_x)=\indeg_N(x)>1$ is a direct consequence of the definition of the $\expand$-operation.
	Hence, in $N'$ there is no vertex $v$ with $\indeg(v)\leq 1$ and $\outdeg(v) = 1$, i.e., $N'$ remains
    phylogenetic.

	Now suppose that $N$ is an $\lca$-network, i.e., that $\lca_N(A)$ is well-defined for every
	non-empty subset $A\subseteq X$. We will now show that $N'$ is also an $\lca$-network. In
	particular, $N$ has the 2-lca property and, as shown above, $N'$ has the 2-lca property. Using
	analogous arguments to those used to show that $\lca_{N'}(x,y)=\lca_N(x,y)$ for all $x,y\in X$ it
	is straightforward to verify that $\lca_{N'}(A)=\lca_N(A)$ for all non-empty subsets $A\subseteq
	X$. Now, consider the vertex $z$ and some subset $A'\subseteq X\cup \{z\}$ that contains $z$. If
	$|A'|\in \{1,2\}$, then the fact that $N'$ has the 2-lca property implies that $\lca_{N'}(A')$ is
	well-defined. Hence, suppose that $|A'|\geq 3$. In this case, there is a subset $A\subseteq X\cap
	A'$ of size $|A|\geq 2$. Since $N$ is an $\lca$-network, $ \lca_N(A)$ is well-defined. In
	particular $A\subseteq X$ and, as previously argued, it holds that $\lca_{N'}(A)=\lca_{N}(A)$.
	Since $|A|>1$, it follows that $\lca_{N'}(A)=u$ for some $u\in V^0(N)$. By construction of $N'$,
	we have $z\prec_{N'}u_x\prec_{N'}u$ for every $x\in A$, so $u$ is a common ancestor of
	$A'=A\cup\{z\}$. Moreover, since no vertex in $V(N')\setminus V(N)$ can be a common ancestor of
	$A$ it follows that every common ancestor $u'$ of $A'$ must satisfy $u'\in V(N)$. Since $A
	\subseteq A'$, every such $u'$ is also a common ancestor of $A$, and $u = \lca_N(A)$ therefore
	implies that $u \preceq_N u'$. Since $u'$ was an arbitrary common ancestor of $A'$, the latter
	arguments imply that $\lca_{N'}(A') = \lca_N(A)$ and that $\lca_{N'}(A')$ is well-defined. In
	conclusion, $N'$ is an $\lca$-network.

	Finally, assume that $N$ has at most $k-1$ hybrid vertices. 
	As argued above, the only vertices that are in $V(N)\cap V(N')$ and that differ in their in- or out-degrees are
	those hybrids $x$ in $N$ that are leaves and for each such hybrid $x$ there is a unique 
	corresponding hybrid vertex $w_x$ in $N'$ added during
	Step~\ref{graft:leafextend} of Definition~\ref{def:graft} by applying $\expand(x)$. 
	In particular, there is a 1-to-1 correspondence
	between hybrid-leaves $x$ and the vertices $w_x$ resulting from $\expand(x)$.
	Each hybrid vertex in $N$ that is not a leaf  will, by construction, remain a hybrid vertex in $N'$. 
	Since $|X|>1$, the added vertex $z$ will serve as a hybrid in $N'$
	that is not contained in $N$.
	The latter arguments imply that the total number of hybrids in $N'$ is at most $k$.  
	This, in particular, implies that every block in $N'$ can have at most $k$ hybrids. 
	Consequently, $N'$ is level-$k$.
\end{proof}

\begin{theorem}\label{thm:N<-v_explains-G}
	Let $G$ be a graph and suppose that $(N,t)$ is a $0/1$-labeled network that explains $G-z$ for some $z\in V(G)$. 
	Then, $(N_{\NWA z}, t_{\NWA z})$ explains $G$ where $t_{\NWA z} \colon V^0(N_{\NWA z}) \to \{0,1\}$ is the map defined
	by putting,  for each	 $w\in V^0(N_{\NWA z})$,
	\begin{equation}\label{eq:mapN-NWA-z}
	t_{\NWA z}(w)\coloneqq\begin{cases}
    t(w) &\text{if } w\in V(N)\\
    1 &\text{if } w=u_x \text{ for some } x\in X \text{ and } \{x,z\}\in E(G)\\
    0 &\text{otherwise, i.e., if either $w=w_x$ or if $w=u_x$ and  $\{x,z\}\notin E(G)$ for some } x\in X 
  \end{cases}
  \end{equation}
\end{theorem}
\begin{proof}
	Let $G$ be a graph and suppose that $(N,t)$ is a $0/1$-labeled network that explains $G-z$ for some
	$z\in V(G)$. Consider now $(N_{\NWA z},t_{\NWA z})$, where $t_{\NWA z}$ is defined as in
	Equation~\eqref{eq:mapN-NWA-z}. Let $x,y\in V(G)$. If $z\notin\{x,y\}$ then, according to
	Equation~\eqref{eq:lca-NWA} in Lemma~\ref{lem:graft-properties}, $\lca_{N_{\NWA z}}(x,y) =
	\lca_{N}(x,y)\eqqcolon w\in V(N)$ and thus, $t_{\NWA z}(w) = t(w)$. 
	Since $(N,t)$ explains $G-z$ and $z\notin\{x,y\}$ we can conclude that 
	$t_{\NWA z}(w) =1$ if and only if $\{x,y\}\in E(G)$. Consider now a vertex $x\in V(G) \setminus\{z\}$ and the vertex $z$. By
	Equation~\eqref{eq:lca-NWA},  $\lca_{N_{\NWA z}}(x,z) = u_x$. By definition of $t_{\NWA z}$, we have 
	$t_{\NWA z}(\lca_{N_{\NWA z}}(x,z)) =1$ if and only if $\{x,y\}\in E(G)$. 
	Consequently, $(N_{\NWA z},t_{\NWA z})$ explains $G$.
\end{proof}

In phylogenetic trees, the number of inner vertices and edges is bounded from above (linearly) by the number
of its leaves. In general, phylogenetic networks lack this property. In fact it is easy to construct
networks on $X$ with $2^{|X|}$ vertices or more. 
Nevertheless, as shown in \cite[Prop.~3.3]{LSH:24}, it is always possible to construct a level-$|X|$ network 
with $O(|X|^2)$ vertices  that can explain a given graph $G=(X,E)$. 
We provide here a slightly stronger statement.

\begin{theorem}\label{thm:every_G_lev-k-ex}
If there is a subset $Y\subsetneq X$ of a graph $G=(X,E)$ such that $G-Y$ is a cograph,
then $G$ is \levIex[$|Y|$] and can be explained by a  $0/1$-labeled phylogenetic lca-network with $O((|Y|+1)|X|)\subseteq O(|X|^2)$ vertices.
In particular, every graph $G=(X,E)$ is \levIex[$k$] for some 
$k<|X|$.
\end{theorem}
\begin{proof}
Let $G=(X,E)$ be a graph and assume that there is a set $Y\subsetneq X$ such that $G-Y$ is a
cograph. If $|Y|=0$, then $G-Y=G$ is a cograph and, by Theorem~\ref{thm:CharCograph}, $G$ is
explained by a phylogenetic $0/1$-labeled tree and thus, by a phylogenetic $0/1$-labeled level-0 network with $O(|X|)$ vertices. 
In particular, $G$ is \levIex[0]. Moreover, any tree is an lca-network \cite{Hellmuth2023}.
Consider now the case when $1\leq|Y|<|X|$. Since $G-Y$ is a cograph,
Theorem~\ref{thm:CharCograph} implies that $G-Y$ is explained by its cotree and thus, 
re-using the previous arguments, by a phylogenetic level-0 lca-network $(N_0,t_0)$. Order the
vertices $Y=\{y_1,\ldots,y_k\}$ arbitrarily, where $k\coloneqq |Y|$. We can, by  repeatedly applying 
Definition~\ref{def:graft}, define a sequence of $k$ digraphs $N_1$, \ldots, $N_k$ by
putting $N_i\coloneqq (N_{i-1})_{\NWA y_i}$ for $i=1,2\ldots, k$. By repeated application 
of Lemma~\ref{lem:graft-properties} and Theorem~\ref{thm:N<-v_explains-G} 
there is a labeling $t_i$ of the level-$i$ network $N_i$ such
that $\mathscr{G}(N_i,t_i)=G-\{y_{i+1},\ldots,y_k\}$. In particular, $(N_k,t_k)$ is a 
phylogenetic level-$k$ lca-network that explains $G-\emptyset=G$, that is, $G$ is \levIex[$k$].
Moreover, when constructing $N_{i+1}$ from $N_i$ we add $O(|X|)$ vertices. 
Hence, the final network $N_k$ has $O(k|X|)$.

To prove the second statement, observe that for every graph $G=(X,E)$ and all sets $Y = X\setminus \{x\}$ for some $x\in X$
it holds that $G-Y$ is a single vertex graph and thus,  a cograph. Reusing the arguments in  preceding paragraph, 
$G$ is \levIex[$k$] for some $k \leq |X|-1$. 
\end{proof}

Finding a subset $Y \subsetneq V(G)$ of minimum size such that $G - Y$ is a cograph is an
NP-hard task \cite{LY:80}. It remains, however, an open question whether the problem of finding
the smallest $k$ such that $G$ is \levIex[$k$] is an NP-hard task as well. There is no direct
relation between the size of a set $Y$ for which $G - Y$ is a cograph and the $k$ for which $G$
is \levIex[$k$]. In particular, the converse of Theorem~\ref{thm:every_G_lev-k-ex} is in general
not satisfied, as illustrated in Figure~\ref{fig:level2counterex}. In this example, $G$ is
\levIex, however, the size of a minimum set $Y$ resulting in a cograph $G - Y$ is $|Y| > 1$. 
A further example is provided in Figure~\ref{fig:level2counterex-primitive}.

\section{Level-1 explainable graphs}
\label{sec:lvl1ex}
In Section~\ref{sec:GeneralDAGs}, we saw that every graph can be explained by some $0/1$-labeled
network. Moreover, only cographs admit an explanation by a $0/1$-labeled tree
and, thus, by a level-$0$ network. Although the number of
vertices in a network $(N,t)$ explaining a graph $G = (X,E)$ can be asymptotically bounded above by
a function of the number $|X|$ of its leaves (cf. Theorem~\ref{thm:every_G_lev-k-ex}), such networks can still be
quite complex. In particular, their level can increase drastically, leading to level-$k$ networks, where $k$ is close to $|X|$, as demonstrated by the construction in
Theorem~\ref{thm:N<-v_explains-G}. This naturally raises the question of whether it is possible to
find a ``simpler'' $0/1$-labeled network that explains $G$. Hence, we explore in this section the
structure of graphs that can be explained by level-1 networks.

To this end, we begin in Section~\ref{subS:MD} by introducing the modular decomposition of graphs.
Every graph admits a unique decomposition, $\MDstrong(G)$, into so-called strong modules. The set
$\MDstrong(G)$ of these modules forms a hierarchy and can therefore be represented as a labeled
tree, the modular decomposition tree $(\MDT_G,\tau_G)$. However, in general, $(\MDT_G,\tau_G)$ is
not $0/1$-labeled, as it uses an additional label ``$\primeL$'' for its vertices. In particular, $G$
is a cograph if and only if $(\MDT_G,\tau_G)$ is $0/1$-labeled. The main goal here is to determine
whether a given graph $G$ is \levIex, and, if so, to provide a level-1 network that explains $G$. To
achieve this, we propose replacing each $\primeL$-labeled vertex with a $0/1$-labeled network. In
Section~\ref{subS:general}, we study the structure of so-called primitive graphs that are \levIex.
It turns out that these graphs are precisely the \emph{near-cographs}, i.e., those graphs for which
the removal of a single vertex results in a cograph. In Section~\ref{subS:primitive}, we combine
these findings to provide several characterizations of \levIex graphs. In particular, this enables us to
to design a linear-time algorithm for recognizing \levIex graphs and constructing $0/1$-labeled
level-1 networks explaining them.

\subsection{Modular decomposition}
\label{subS:MD}

We begin by providing the necessary definitions for characterizing graphs that are \levIex. To
recall, the class of cographs is precisely the class of graphs that can be explained by a
$0/1$-labeled tree. However, not all graphs are cographs. Once $G$ contains an induced $P_4$, there
is no $0/1$-labeled tree $(T,t)$ that can explain $G$. Nevertheless, there is a way to construct a
labeled tree, called the \emph{modular decomposition tree}, which captures at least partial
information about the structure of $G$.

A \emph{module} $M$ of a graph $G=(X,E)$ is a non-empty subset $M\subseteq X$ such that for all
$x,y\in M$ it holds that $N_G(x)\setminus M = N_G(y)\setminus M$, where $N_G(x)$ denotes the set of
all vertices in $G$ adjacent to $x$. In other words, $M$ is a module if, for all $x,y\in M$ and all
$z\notin M$, $x$ is adjacent to $z$ if and only if $y$ is adjacent to $z$. Note that $X$ and each
singleton set $\{x\}$ for $x\in X$ are modules of $G$. They are called the \emph{trivial} modules of
$G$. All other modules are \emph{non-trivial}. The set $\MD(G)$ comprising all modules of $G$ in
particular contains the trivial modules of $G$ and is therefore always non-empty. A graph $G$ is
called \emph{primitive} if it has at least four vertices and $\MD(G)$ contains trivial modules only.
The smallest primitive graph is the path $P_4$ on four vertices. Moreover, every primitive graph
contains an induced $P_4$ (c.f. \cite[L.~1.5]{Sumner:1973}) and is therefore not a cograph.
Modules and modular decomposition have appeared under many different names, we refer to the surveys
\cite{Habib:2010,Kelly:1985} for recollections of the history of these concepts; in particular note
that modules have also appeared under the names \emph{closed sets} \cite{Gallai:1967}, \emph{clumps}
\cite{Aschbacher:1976, Blass:1978}, \emph{autonomous sets} \cite{Mohring:1984,Mohring:1985} and
\emph{clans} \cite{Ehrenfeucht:1990A,Ehrenfeucht:1990B,Ehrenfeucht:1990C,Ehrenfeucht:1994}.
Similarly, primitive graphs \cite{Ehrenfeucht:1990A,Ehrenfeucht:1990B,Ehrenfeucht:1990C} are also
known as \emph{prime} graphs \cite{Cournier:1994, Habib:2010} and \emph{indecomposable} graphs
\cite{Cournier:1998,Ille:1997,Schmerl:1993,Sumner:1973}.

A module $M$ of $G=(X,E)$ is \emph{strong} if $M$ does not overlap with any other module of $G$. 
Clearly, all trivial modules of $G$ are strong. The
set of strong modules $\MDstrong(G)\subseteq \MD(G)$ is uniquely determined
(see e.g. \cite{Ehrenfeucht:1994}) and $\MDstrong(G)$ is a hierarchy on $X$.
One can furthermore distinguish between three types of strong modules; a module $M\in\MDstrong(G)$ is:
\begin{center}
\begin{itemize}[noitemsep]
	\item \emph{series} if $\overline{G[M]}$ is disconnected,\smallskip
	\item \emph{parallel} if $G[M]$ is disconnected and \smallskip
	\item \emph{prime} otherwise, i.e., if both $\overline{G[M]}$ and $G[M]$ are connected.
\end{itemize}
\end{center}
We can in particular study the strong modules of $G=(X,E)$ by considering the \emph{modular decomposition tree} (MDT) of $G$,
that is, the Hasse diagram $\hasse(\MDstrong(G))$. More precisely, we can equip $\MDT_G\doteq \hasse(\MDstrong(G))$ with a labeling
$\tau_G:V^0(\MDT_G)\to\{0,1,\primeL\}$ defined by setting 
\[\tau_G(M)\coloneqq \begin{cases}
	0 & \text{if $M$ is a parallel module}  \\
	1 & \text{if $M$ is a series module}  \\
	\primeL & \text{if $M$ is a prime module}
\end{cases}
\]
for all $M\in V^0(\MDT_G)=\MDstrong(G)\setminus\{\{x\}\,:\,x\in X\}$. Since $\MDstrong(G)$ is a
hierarchy, $\MDT_G$ is a phylogenetic tree (c.f.\ Theorem~\ref{thm:CharTreeCluster}). 
Hence,  $(\MDT_G, \tau_G)$ is a natural generalization of cotrees that, however, does 
in general not explain $G$. Although, for all distinct $x,y\in L(\MDT_G)=V(G)=X$
it holds that $\tau_G(\lca_{\MDT_G}(x,y))=1$ implies $\{x,y\}\in E(G)$ and 
$\tau_G(\lca_{\MDT_G}(x,y))=0$ implies that  
$\{x,y\}\notin E(G)$, the converse of these two implications is, in general, not satisfied. 
In particular, by Theorem~\ref{thm:CharCograph} and the definition of cotrees, 
$G$ is explained by $(\MDT_G, \tau_G)$ precisely if $G$ is a cograph and thus, if 
 $(\MDT_G, \tau_G)$ does not contain inner vertices $v$ with $\tau_G(v)=\primeL$. 

Since  $\MDstrong(G)$ is uniquely determined for all graphs $G=(X,E)$ and since it does not contain overlapping modules,
there is a unique partition $\Mmax(G) = \{M_1,\dots, M_k\}$ of $X$
into $k\geq 2$ inclusion-maximal strong modules $M_j\neq X$ of $G$, whenever $G$ contains $|X|\geq 2$  vertices \cite{Ehrenfeucht:1990A,Ehrenfeucht:1990B}.

A well-known fact, that is easy to prove, is the following
\begin{observation}\label{obs:parallel-series}
	Let $G$ be a graph, $(\MDT,\tau)$ be its MDT and $M$ be a  strong module of $G$ such that $|M|>1$.  
	Then $\Mmax(G[M]) = \{M_1,\dots, M_k\}$ contains $k>1$ elements. Moreover, 
	If $\tau(M) = 0$, then $G[M] = G[M_1]\union G[M_2] \union \dots \union G[M_k]$
	and if $\tau(M) = 1$, then $G[M] = G[M_1]\join G[M_2] \join \dots \join G[M_k]$
\end{observation}

For strong modules $M$ such that $\tau(M)=\primeL$, the situation is more complicated, but can be approached as follows.
Two disjoint modules $M, M'\in \MD(G)$ are \emph{adjacent} if each
vertex of $M$ is adjacent to all vertices of $M'$ in $G$; the modules are
\emph{non-adjacent} if none of the vertices of $M$ is adjacent to a vertex of
$M'$ in $G$. By definition of modules, every two disjoint modules $M, M'\in \MD(G)$ are
either adjacent or non-adjacent \cite[Lemma 4.11]{Ehrenfeucht:1990A}. One can therefore
define the \emph{quotient graph} $G/\Mmax(G)$ which has $\Mmax(G)$ as its vertex set
and $\{M_i,M_j\}\in E(G/\Mmax(G))$ if and only if $M_i$ and $M_j$ are adjacent
in $G$. For later reference, we provide the following simple result.

\begin{figure}[t]
	\centering
	\includegraphics[width=0.75\textwidth]{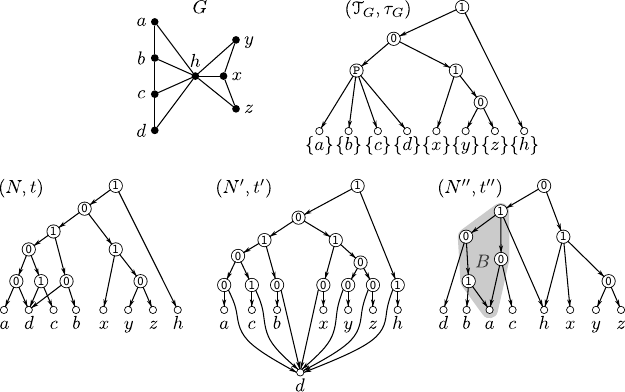}
	\caption{
	Shown is a graph $G$ that is explained by several $0/1$-labeled level-1
	networks: $(N,t)$, $(N',t')$, and $(N'',t'')$. Note that $G$ is a near-cograph as $G-d$ is a cograph. 
	Here, $(N,t)$ is a pvr-network obtained from the MDT $(\MDT_G,\tau_G)$ of $G$ by replacing
	the $\primeL (\mathtt P)$-labeled vertex by a $0/1$-labeled level-1 network according to Definition~\ref{def:pvr}. Moreover, 
	according to Definition~\ref{def:graft} and Equation~\eqref{eq:mapN-NWA-z},
	$(N',t') = (\tilde{T}_{\NWA d}, \tilde{t}_{\NWA d})$ with $(\tilde{T}, \tilde{t})$ being the cotree of $G-d$.	 \newline
	In contrast, the network $(N'',t'')$ is not obtained from the MDT of $G$ resp. cotree of $G-z$ 
	in the manner as the other two networks.
	Since the cluster $\{y,z\} \in \mathfrak{C}_{N''}$ overlaps no cluster in $\mathfrak{C}_{N''}$,
	Corollary~\ref{cor:nonoverlap-module} implies that $\{y,z\}$ is a module of $G$. The same is true
	for the set $\{x,y,z\}$, which is a (strong) module obtained as the union of clusters $\{x\} 
	\cup \{y,z\}$, which does not overlap with any cluster in $\mathfrak{C}_{N''}$.\newline
	According to Lemma~\ref{lem:block-module}, another module of $G$ is given by $\LL(B) =
	\{a,b,c,d\} \subsetneq \CC_{N''}(\max_B)$ for the non-trivial block $B$ of $N''$ (highlighted in gray-shaded area). 
	Note that $\CC_{N''}(\max_B) =
	\{a,b,c,d,h\} \in \mathfrak{C}_{N''}$ is not a module of $G$ as $h$ is adjacent to $x$ but 
	none of the vertices $a,b,c,d$ is adjacent to $x$.
	}
	\label{fig:blocks-modules}
\end{figure}

\begin{observation}[\cite{Habib:2010}]\label{obs:quotient} 
The quotient graph $G/\Mmax(G)$ with $\Mmax(G) = \{M_1 , \dots , M_k\}$ is
isomorphic to any subgraph induced by a set $W\subseteq V$ such that $|M_i \cap
W | = 1$ for all $i \in \{1, \dots,k\}$.
\end{observation}

On a related note, the following lemma provides additional insight into the structure 
of induced primitive subgraphs in relation to the strong modules of $G$.

\begin{lemma}[{\cite[L.~3.4]{fritz2020cograph}, \cite[L.~2.5]{HS:24} \& \cite[Thm.~2.17]{Engelfriet:1996}}]\label{lem:Hprimitive-in-primeM}
 Let $H$ be an induced primitive	 subgraph of $G$ and $M\in \MDstrong(G)$ be a
 strong module that is inclusion-minimal w.r.t.\ $V(H)\subseteq M$. Then, $M$ is
 prime and $H$ is isomorphic to an induced subgraph of $G[M]/\Mmax(G[M])$.
 In particular, $G[M]/\Mmax(G[M])$ is primitive for all non-trivial prime modules $M$ of $G$.
 Moreover, every prime module is strong.
\end{lemma}

In the following, we explore the relationships between modules in graphs and clusters in $0/1$-labeled networks that explain them.

\begin{lemma}\label{lem:nonoverlap-module-stronger}
	Let $(N,t)$ be a $0/1$-labeled network with the 2-lca property and consider some non-empty subset $\mathfrak{C}'\subseteq\mathfrak{C}_N$. If
	$M\coloneqq\cup_{C\in\mathfrak{C}'}C$ overlaps with no cluster in $\mathfrak{C}_N$, then $M$ is a module of $\mathscr{G}(N,t)$.
\end{lemma}
\begin{proof}
	Let $(N,t)$ be a $0/1$-labeled network on $X$ with the 2-lca property. Hence
	$G\coloneqq\mathscr{G}(N,t)$ is well-defined. Assume that
	$\mathfrak{C}'\subseteq\mathfrak{C}_N$ is a nonempty collection of clusters such that
	$M\coloneqq\bigcup_{C\in\mathfrak{C}'} C$ overlaps with no cluster in $\mathfrak{C}_N$. If
	$| M|\in\{1,|X|\}$, then $ M$ is a trivial module of $G$. Thus, assume henceforth that
	$1<| M|<|X|$. In particular, there exists a vertex $x\in X\setminus  M$ and two distinct
	vertices $y,y'\in M$. Moreover, since $N$ has the 2-lca property, the vertices
	$u\coloneqq\lca_N(x,y)$ and $u'\coloneqq\lca_N(x,y')$ are well-defined. Since $x,y\in\CC_N(u)$,
	and $y\in M$ and $x\notin M$, the fact that $ M$ does not overlap with $\CC_N(u)$ implies
	that $ M\subseteq \CC_N(u)$. Since $y'\in M$, it additionally follows that
	$\{x,y'\}\subseteq\CC_N(u)$. Therefore Lemma~\ref{lem:lca-cluster} implies that
	$\lca_N(x,y')\preceq_N u$ i.e. $u'\preceq_N u$. By analogous arguments, one shows that
	$u\preceq_N u'$. Hence, $u=u'$. Since $x\in X\setminus  M$ and $y,y'\in M$ have been
	arbitrarily chosen it follows that $\lca_N(x,y)=\lca_N(x,y')$ and, therefore,
	$t(\lca_N(x,y))=t(\lca_N(x,y'))$ for all $y,y'\in M$ and all $x\notin M$. This and the
	fact that $(N,t)$ explains $G$, implies that $ M$ is a module of $G$.
\end{proof}

	As a simple consequence of Lemma~\ref{lem:nonoverlap-module-stronger}, we obtain
\begin{corollary}\label{cor:nonoverlap-module}
	Let $(N,t)$ be a $0/1$-labeled network with the 2-lca property. If there is a vertex $v$ such
	that $\CC_N(v)$ overlaps with no cluster in $\mathfrak{C}_N$, then $\CC_N(v)$ is a module of
	$\mathscr{G}(N,t)$.
\end{corollary}

Lemma~\ref{lem:nonoverlap-module-stronger} and Corollary~\ref{cor:nonoverlap-module} are illustrated in Figure~\ref{fig:blocks-modules}.
In general, $\MDstrong(G)$ is a proper subset of $\MD(G)$, that is, $G$ can contain overlapping
modules. In particular, the module $M$ defined in Lemma~\ref{lem:nonoverlap-module-stronger} is
not necessarily strong. As an example, consider the network $(N'',t'')$ shown in
Figure~\ref{fig:level2counterex} where any union of clusters $\cup_{i\in
I}\{a_i,b_i,c_i,d_i\}$ forms a module in $G = \mathscr{G}(N,t)$. However, for example,
$\{a_1,b_1,c_1,d_1,a_2,b_2,c_2,d_2 \}$ and $\{a_2,b_2,c_2,d_2,a_3,b_3,c_3,d_3\}$ overlap and are
therefore not strong.

Note that the converse of Lemma~\ref{lem:nonoverlap-module-stronger} is, in general, not satisfied.
That is, even though every module $M$ of $G = \mathscr{G}(N,t)$ can be expressed as the union of clusters (namely, as $M=\cup_{x\in M}\{x\}$)
 there may exist a cluster $C\in \mathscr{C}_N$ such that $M$ and $C$ overlap.
 As an example, consider the graph $G = \mathscr{G}(N'',t'')$ and the
modular decomposition tree $(\MDT_{G}, \tau_{G})$ of $G$ as shown in Figure~\ref{fig:blocks-modules}. Here,
$M \coloneqq \{a,b,c,d,x,y,z\}$ corresponds to a vertex in $(\MDT_{G}, \tau_{G})$, namely the $0$-labeled
child of $\rho_{\MDT_{G}}$. Thus, $M$ is a strong module. However, $M$ overlaps with the cluster
$\CC_{N''}(\max_B) = \{a,b,c,d,h\}$ in $N''$. 

We can obtain additional modules of $G = \mathscr{G}(N,t)$ by considering the union of leaf-descendants of the vertices $v\neq \max_B$ in a non-trivial block $B$, as follows.
Let $B$ be a nontrivial block of a network $N$. By Lemma~\ref{lem:unique-maxB}, $B$ has a unique $\preceq_N$-maximal element $\max_B$. Put
\[\LL(B)\coloneqq\{x\in X\,:\, \text{ there is some } v\in V(B)\setminus\{ \mathrm{max}_B \} \text{ s.t. } x\in\CC_N(v)\}=\bigcup_{v\in V(B)\setminus\{\max_B\}}\CC_N(v).\]
By Lemma~\ref{lem:lca-cluster}, $\LL(B)\subseteq\CC_N(\max_B)$ holds but, as shown in Figure~\ref{fig:blocks-modules}, there are cases when $\LL(B)\neq\CC_N(\max_B)$.

\begin{lemma}\label{lem:block-module}
  For every nontrivial block $B$ of a $0/1$-labeled network $(N,t)$ with the 2-lca property, 
  the set $\LL(B)$ is a module of $\mathscr{G}(N,t)$. 
  If, additionally, $N$ is 2-lca-relevant then
  $|\LL(B)|>1$.
\end{lemma}
\begin{proof}
	Let $(N,t)$ be a $0/1$-labeled network on $X$ with the 2-lca property. Hence
	$G\coloneqq\mathscr{G}(N,t)$ is well-defined. By definition, $V(G)=X$ holds. Now consider any
	nontrivial block $B$ of $N$. By Lemma~\ref{lem:unique-maxB}, $B$ has a unique $\preceq_N$-maximal
	element $\max_B$ and, by Lemma~\ref{lem:lca-cluster}, $\LL(B)\subseteq\CC_N(\max_B)$ holds. 

	Assume, for contradiction, that $\LL(B)$ overlaps with some $C\in \mathfrak{C}_N$ and let $v\in
	V(N)$ be a vertex that satisfies $\CC_N(v) =C$. In this case, $\LL(B)\cap \CC_N(v)\neq \emptyset$
	implies that there is a vertex $w\in V(B)\setminus \{\max_B\}$ such that $\CC_N(w) \cap
	\CC_N(v)\neq \emptyset$. We distinguish now between the cases that $v$ and $w$ are
	$\preceq_N$-incomparable or not. Assume first that $v$ and $w$ are $\preceq_N$-incomparable.
	Lemma~\ref{lem:unique-maxB} implies that $w$ and $v$ are located in a common block $B'$. Observe
	that $w\neq \max_{B'}$ since, otherwise, $v\preceq_N w$ would hold. Hence, $w\notin
	\{\max_B,\max_{B'}\}$ and Lemma~\ref{lem:unique-maxB} implies that $B=B'$. Now there are two
	cases for $v$; either $v=\max_B$ or not. If $v=\max_B$, then $w\preceq_N v$; a contradiction. If
	$v\ne\max_B$, then, by definition, $\CC_N(v)\subseteq \LL(B)$ and $\LL(B)$ does not overlap with
	$\CC_N(v)$; a contradiction. Hence, the case that $v$ and $w$ are $\preceq_N$-incomparable cannot
	occur.

	Assume now that $v$ and $w$ are $\preceq_N$-comparable. If $v\preceq_N w$,
	Lemma~\ref{lem:lca-cluster} implies that $\CC_N(v)\subseteq \CC_N(w)\subseteq \LL(B)$;
	contradicting the fact that $\CC_N(v)$ and $\LL(B)$ overlap. Hence, it must hold that $w\prec_N
	v$. If $\max_B \preceq_N v$, Lemma~\ref{lem:lca-cluster} implies that we have $\LL(B) \subseteq
	\CC_N(\max_B) \subseteq \CC_N(v)$; again a contradiction. If $v \prec_N \max_B$, there is a
	directed path from $\max_B$ to $w$ that contains $v$ and each such path is, by
	Lemma~\ref{lem:unique-maxB}, completely contained in $B$. Thus, $v\in V(B)$. But then $v\neq
	\max_B$ implies that $\CC_N(v)\subseteq \LL(B)$; again a contradiction. 
	Therefore, $v$ and $\max_B$ must be $\preceq_N$-incomparable. Since 
	$\LL(B) \subseteq	\CC_N(\max_B)$ it follows that $\CC_N(v)\cap \CC_N(\max_B)\neq \emptyset$ and,
	by Lemma~\ref{lem:unique-maxB},
	$v$ and $\max_B$ are contained in some common non-trivial block $B'$. 
	Hence, there is a directed path $P_{\max_{B'}v}$ from $\max_{B'}$ to $v$
	and a directed path $P_{\max_{B'}\max_B}$ from $\max_{B'}$ to $\max_B$. Note that $P_{\max_{B'}v}$
	does not contain $\max_B$ and that $P_{\max_{B'}\max_B}$ does not contain $v$, since 
	$v$ and $\max_B$ are $\preceq_N$-incomparable. Let now $w''$ be the 
	$\preceq_N$-minimal vertex
	that is contained in $P_{\max_{B'}v}$ and $P_{\max_{B'}\max_B}$ and denote with $P_{w''v}$ and $P_{w''\max_B}$
	the subpath of $P_{\max_{B'}v}$ from $w''$ to $v$  and the subpath of $P_{\max_{B'}\max_B}$ from $w''$ to $\max_B$, respectively. 	
	By construction, $P_{w''v}$ and $P_{w''\max_B}$ have only the vertex $w''$ in common.	
	Since $w\prec_N v$ and $w\prec_N \max_B$, there is a directed $vw$-path $P_{vw}$	
	and directed $\max_B w$-path $P_{\max_Bw}$. Let $w'$ be the $\preceq_N$-maximal vertex
	that is contained in $P_{vw}$ and $P_{\max_Bw}$ and denote with $P_{vw'}$ and $P_{\max_B w'}$
	the subpath of  $P_{vw}$ from $v$ to $w'$  and the subpath of $P_{\max_Bw}$ from $\max_B$ to $w'$, respectively.
	By construction, $P_{vw'}$ and $P_{\max_B w'}$ have only the vertex $w'$ in common.
	Since $w'\prec_N v$ and $v$ and $\max_B$ are $\preceq_N$-incomparable, we have $w'\neq \max_B$. In particular,
	$w'\in V(B)\setminus\{\max_B\}$, since $w'$ is a vertex of the path $P_{\max_Bw}$ which, by Lemma~\ref{lem:unique-maxB}, is completely contained in $B$.
	Since $N$ is a DAG, we can now construct a directed path $P$ by combining $P_{w''v}$ with $P_{vw'}$
	and a directed path $P'$ by combining $P_{w''\max_B}$ with $P_{\max_Bw'}$ to obtain
	two directed paths that have only the vertices $w'$ and $w''$ in common. In particular, 
	the subgraph of $N$ induced by the vertices of $P$ and $P'$ is biconnected and is thus contained
	in some non-trivial block $B''$. By construction, $B''$ contains $v, w'$ and $w''$.
	This and  $w'\prec_N w''$  implies that $w'\in V(B'')\setminus \{\max_{B''}\}$. As argued above, 
	$w'\in V(B)\setminus\{\max_B\}$. Thus,
	we can apply  Lemma~\ref{lem:unique-maxB} to conclude that $B=B''$. Since $v\in V(B'')$, we have $v\in V(B)$.
	Therefore, $v\preceq_N \max_B$ must hold; a contradiction.  In summary,	 
	$\LL(B)$ cannot overlap with any cluster
	$C\in \mathfrak{C}_N$. By Lemma~\ref{lem:nonoverlap-module-stronger}, $\LL(B)$ is a
	module of $\mathscr{G}(N,t)$.

	For the second statement, furthermore assume that $N$ has the 2-lca property and is 2-lca-relevant. 
	Since $B$  is a non-trivial block, there exist vertices $w,v\in V(B)$ such that $w\prec_N v\prec_N \max_B$.
	Moreover, Lemma \ref{lem:lca-cluster} implies that $\CC_N(w)\subseteq\CC_N(v)$. 
	Since $N$ is 2-lca-relevant, $v$ is a least common ancestor for some subset $A\subseteq X$ of size $|A|\leq 2$.
	Thus,  $\CC_N(w) = \CC_N(v)$ is not possible as, otherwise,  $w\prec_N v$ implies that 
	$v$ is not a  least common ancestor for any subset $A\subseteq X$. 
	Hence, $\CC_N(w) \subsetneq \CC_N(v)$ holds. Therefore, $|\CC_N(v)|>1$ which, together with $\CC_N(v)\subseteq \LL(B)$, 
	implies that $|\LL(B)|>1$ 
\end{proof}

\subsection{Characterizing primitive {\levIex} graphs}
\label{subS:primitive}

\begin{figure}
  \centering
  \includegraphics[width=0.6\textwidth]{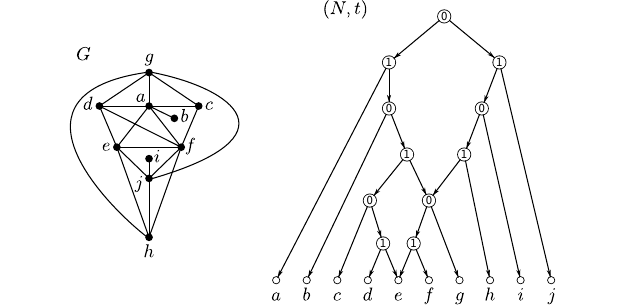}
  \caption{	A primitive graph $G=\mathscr{G}(N,t)$ that is explained by the $0/1$-labeled level-2 network $(N,t)$. 
			  Here, there is no subset $Y\subseteq V(G)$ of size $|Y|\leq 2$ such that $G-Y$ is a cograph. 
  			  To see this, assume, for contradiction, that there is a set $Y$ of size $|Y|\leq 2$ resulting in the  cograph  $G-Y$.
  			  First observe that there are two intertwined induced $P_4$s $G[\{b,a,f,j\}]$ and $G[\{a,f,j,i\}]$ that are both vertex-disjoint with 
  			  the induced $P_4$ $G[\{c,g,d,e\}]$. Hence, $Y$ must contain one vertex of $a,f,j$ and one  vertex of   			  
  			  $c,g,d,e$ to destroy these $P_4$. If $j\in Y$, then $Y = \{j,v\}$ with $v\in \{c,g,d,e\}$ 
  			  would leave the induced $P_4$ $G[\{b,a,f,h\}]$ in $G-Y$. Hence, $j\notin Y$. 
  			  If $f\in Y$ and $Y = \{f,e\}$, then the induced $P_4$ $G[\{b,a,g,j\}]$ is in $G-Y$.
  			  Thus, $Y\neq \{f,e\}$ and, if $f\in Y$ then, $Y = \{f,v\}$ with $v\in \{c,g,d\}$ must hold
  			  in which case $G-Y$ contains the induced $P_4$ $G[\{a,e,j,i\}]$. 
  			  Thus, $f\in Y$ is not possible. Thus, $a\in Y$ must hold. 
  			  If $Y=\{a,g\}$ or $Y=\{a,c\}$, the induced $P_4$ $G[\{d,e,j,i\}]$
  			  remains in $G-Y$. If $Y=\{a,d\}$ or $Y=\{a,e\}$,
  			  the induced $P_4$ $G[\{c,f,j,i\}]$  remains in $G-Y$.
  			  Thus, none of the combinations $Y=\{v,w\}$ with $v\in \{a,f,j\}$
  			  $w\in \{c,g,d,e\}$ yield a cograph. Thus, there is no set $Y$ of size $|Y|\leq 2$ resulting in the  cograph  $G-Y$.
			}
  \label{fig:level2counterex-primitive}
\end{figure}

We explore now the structure of primitive \levIex graphs. First, we show that any primitive
\levIex graph is explained by a $0/1$-labeled level-1 network with fairly constrained structural
properties. We then prove that primitive \levIex graphs are precisely the primitive
near-cographs, as defined below.

\begin{lemma}\label{lem:lvl1+primitive=>unique-hybrid}
	If $G$ is a primitive {\levIex} graph, then there exists a $0/1$-labeled phylogenetic level-1 network $(N,t)$ that explains $G$ 
	and that contains exactly one hybrid and this hybrid is a leaf. In particular, $N$ contains precisely
	one non-trivial block.
\end{lemma}
\begin{proof}
	Let $G$ be a primitive graph and assume that $(N,t)$ is a $0/1$-labeled level-1 network that explains
	$G$. By definition, $N$ has the 2-lca property. Moreover,
	by Theorem~\ref{thm:regular-version}, we can assume w.l.o.g.\ that $N$ is a 
	2-lca-relevant and regular level-1 network. By Lemma~\ref{lem:regular-props}, 
	$N$ is phylogenetic.
	Since $G$ is primitive, it cannot be a cograph and
	Theorem~\ref{thm:CharCograph} implies that $G$ cannot be explained by a tree. Thus, $N$ must
	contain at least one non-trivial block $B$. By Lemma~\ref{lem:block-module}, the set
	$\LL(B)$ is a module of $G$ and $|\LL(B)|>1$. Since $G$ is primitive, we must
	thus have that $\LL(B)=X$. Moreover, Lemma~\ref{lem:unique-maxB} implies that
	$v\preceq_N \max_B$ for all $v\in V(B)$. This together with Lemma~\ref{lem:lca-cluster} implies
	that $\LL(B)\subseteq\CC_N(\max_B)$. Hence, $\CC_N(\max_B)=X$. Since
	$\CC_N(\rho_N)=X$ and every vertex of a regular network is associated with a unique cluster (cf.\
	Lemma~\ref{lem:regular-props}), it follows that $\max_B=\rho_N$.

	Assume, for contradiction, that there is a non-trivial block $B'$ of $N$ such that $B'\neq B$. By
	applying the same arguments as for $B$, we can conclude that $\LL(B')=X$ and
	$\max_{B'}=\rho_N$. Let $x\in X$. Thus, there are vertices $u\in V(B)\setminus\{\max_{B}\}$
	and $v\in V(B')\setminus\{\max_{B'}\}$ such that $x\in \CC_N(u)\cap \CC_N(v)$. Observe first that
	$u\preceq_N v$ is not possible since, otherwise, there is a directed $\rho_Nu$-path in $N$ containing $v$
	and each such path is by Lemma~\ref{lem:unique-maxB} completely contained in $V(B)$, which
	contradicts $v\notin V(B)$. Similarly, $v\preceq_N u$ cannot occur. Hence, $u$ and $v$ are
	$\preceq_N$-incomparable. Lemma~\ref{lem:unique-maxB} implies that $u$ and $v$ are contained in a
	common non-trivial block; a contradiction to $B\neq B'$. Hence, there is no 
	non-trivial block $B'$ of $N$ such that $B'\neq B$.	
	In particular, all hybrid vertices of $N$ belong to some non-trivial block and since there is only one,
	namely  $B$, all hybrid vertices of $N$ are vertices of $B$. 
	Since $N$ is level-1 there is thus a unique hybrid vertex $h$ in $N$.

  Consider now the unique hybrid $h$ and its cluster $\CC_N(h)$. Assume first that there exists a
  vertex $v$ in $N$ such that $\CC_N(v)$ and $\CC_N(h)$ overlap. Then, by \cite[L.~18 \&
  L.~19]{Hellmuth2023}, there is a hybrid vertex $h'$ in $B$ such that $h'\prec_N h$;
  contradicting the uniqueness of $h$. Hence, $\CC_N(h)$ does not overlap
  with any cluster in $\mathfrak{C}_N$. This and Corollary~\ref{cor:nonoverlap-module} implies that
  $\CC_N(h)$ is a module of $G$. Since $G$ is primitive, $\CC_N(h)$ must be a trivial module.
  Since $h\neq \rho_N$, Lemma~\ref{lem:regular-props} ensures that $\CC_N(h)\neq\CC_N(\rho_N)=X$.
  Thus, $\CC_N(h)=\{x\}$ for some $x\in X$. Since $\CC_N(h) = \{x\}= \CC_N(x)$ and due to Lemma~\ref{lem:regular-props}, 
  we can conclude that $h=x$ i.e. that $h$ is a leaf.  
\end{proof}

\begin{definition}
	A graph $G=(V,E)$ is a \emph{near-cograph} if $|V|=1$ or if there is a vertex $v\in V$ such that $G-v$ is a cograph.
\end{definition}

By Theorem~\ref{thm:CharCograph}, 
the property of being a cograph is hereditary. Hence, every cograph is, in particular, a near-cograph. In addition, a cograph does not 
 contain induced $P_4$s. Thus, we obtain
\begin{observation}\label{obs:allP4s}
A graph $G$ is near-cograph if and only if 
$G$ contains a vertex that is located on all induced $P_4$s in $G$ if there are any.  
\end{observation}

Next, we provide a simple characterization of near-cographs that, in particular, shows that
near-cographs form a hereditary graph class that is closed under complementation.
\begin{lemma}\label{lem:char-near-cograph}
	The following statements are equivalent. 
	\begin{enumerate}[label=(\arabic*)]
		\item $G$ is a near-cograph.
		\item $\overline{G}$ is a near-cograph.
		\item Every induced subgraph of $G$ is a near-cograph.
	\end{enumerate}
\end{lemma}
\begin{proof} The equivalences are trivial for single-vertex graphs. Hence, suppose that 
$G = (V,E)$ is a graph with $|V|>1$.
If $G$ is a near-cograph, then $G-v$ is a cograph for some $v\in V(G)=V(\overline G)$.
By Theorem~\ref{thm:CharCograph}, the complement of a cograph is a cograph and it follows that $\overline{G-v} = \overline{G}-v$
is a cograph. Hence, $\overline G$ is a near-cograph. 
Similar arguments show that $G$ is a near-cograph whenever 
$\overline{G}$ is, i.e., (1) and (2) are equivalent. 

Suppose now that $G$ is a near-cograph. Assume, for contradiction, that 
$G$ contains an induced subgraph $H$ that is not a near-cograph. 
By Observation~\ref{obs:allP4s}, $H$ contains no vertex that is located on all induced $P_4$s. 
Since $H$ is an induced subgraph of $G$, this property remains in $G$ and Observation~\ref{obs:allP4s} implies
that $G$ is not a near-cograph; a contradiction. Hence, (1) implies (3).
Moreover, (3) implies (1) since $H=G$ is an induced subgraph of $G$ and thus, a near-cograph. 
\end{proof}

As an immediate consequence of Theorem~\ref{thm:every_G_lev-k-ex} and the definition of
near-cographs, we also obtain the following
\begin{corollary}\label{cor:near-cograph=>lev1ex}
	Every near-cograph is \levIex.
\end{corollary}
The converse of Corollary~\ref{cor:near-cograph=>lev1ex} does not hold; consider, for example, the
\levIex graph $G$ in Figure~\ref{fig:level2counterex}, consisting of $k > 1$ vertex-disjoint $P_4$s.
By Observation~\ref{obs:allP4s}, this implies that $G$ is not a near-cograph. Despite this, we are
now in a position to provide a characterization of primitive \levIex graphs.

\begin{theorem}\label{thm:char-primitive-level1exp}
  Let $G$ be a primitive graph. The following statements are equivalent:
  \begin{enumerate}[label=(\alph*)]
    \item $G$ is a near-cograph.
    \item $G$ is \levIex.
  \end{enumerate}
\end{theorem}
\begin{proof}
	Let $G=(X,E)$ be a primitive graph. First assume that $G$ is a near-cograph. Hence, there is a
	vertex $v\in X$ such that $G-v$ is a cograph. 
	By Theorem~\ref{thm:every_G_lev-k-ex}, $G$ is \levIex[$|\{v\}|$], that is, \levIex.
	
	Conversely, assume that $G$ is \levIex. Since $G$ additionally is primitive,
	Lemma~\ref{lem:lvl1+primitive=>unique-hybrid} implies that there exists a $0/1$-labeled phylogenetic level-1
	network $(N,t)$ on $X$ that explains $G$ such that $N$ contains exactly one hybrid $h\in X$
	and exactly one non-trivial block $B$. By definition of primitivity, 
	$|X|\geq 4$ and thus, $X\setminus \{h\}\neq \emptyset$.

	We show now that  $\mathfrak{C}'\coloneqq \mathfrak{C}_N\cap (X\setminus \{h\})$ is a hierarchy. 
	Observe first that $X\setminus \{h\}\in \mathfrak{C}'$ and $\{x\} \in \mathfrak{C}'$ for all $x\in X\setminus \{h\}$, 
	which ensures that $\mathfrak{C}'$ is a clustering system. 
	Assume, for contradiction,
	that there are overlapping clusters $A, B \in \mathfrak{C}'$ and put $C \coloneqq A\cap B$. 
	The clusters $A$ and $B$ stem from two clusters $\CC_N(u), \CC_N(v) \in \mathfrak{C}$ via $A = \CC_N(u)\setminus \{h\}$ and $B = \CC_N(v)\setminus \{h\}$. 
	Note that $A = \CC_N(u)$ or $B = \CC_N(v)$ may be possible. 
	Since $A\cap B = C\neq \emptyset$ and $h\notin C$ it follows that 
	$\CC_N(u)\cap \CC_N(v)\neq \emptyset$ and $\CC_N(u)\cap \CC_N(v)\neq \{h\}$. 	
	Since $h$ is the only hybrid vertex in $N$, 
	Lemma~\ref{lem:lvl1-intersections} implies that 
	$\CC_N(v)\subseteq \CC_N(u)$ or $\CC_N(u)\subseteq \CC_N(v)$ holds.
	We may assume w.l.o.g.\ that $\CC_N(u)\subseteq \CC_N(v)$. 
	If $h\in \CC_N(u)$, then $h\in \CC_N(v)$ and we obtain $A\subseteq B$; a contradiction. 
	If $h\notin \CC_N(u)$, then $A=\CC_N(u) \subseteq B$; again a contradiction. Thus, there are no overlapping 
	$A, B \in \mathfrak{C}'$. Consequently, $ \mathfrak{C}'$ is a hierarchy.
	
	Since $\mathfrak{C}'$ is a hierarchy, Theorem~\ref{thm:CharTreeCluster} ensures that $\hasse(\mathfrak{C}')$ is a phylogenetic tree. 
	By Proposition~\ref{prop:lcaN-explain}, $T\doteq\hasse(\mathfrak{C}')$ can be equipped with 
	a $0/1$-labeling $t'$ such that $(T,t')$ explains $G[X\setminus \{h\}] = G-h$. 
	By Theorem~\ref{thm:CharCograph}, $G-h$ is a cograph and, therefore, $G$ is a near-cograph.
\end{proof}

	 An alternative proof strategy for the ``(b) $\Rightarrow$ (a)'' direction could
    be to consider the network $(N,t)$ on $X$ that explains $G=(X,E)$, where $N$ contains exactly one
    hybrid $h \in X$ and exactly one non-trivial block $B$. By removing $h$ and its incident edges
    from $N$, we obtain a $0/1$-labeled tree $(T,t')$ on $X\setminus \{h\}$. 
    It then suffices to show that $(T,t')$
    explains $G-h$ to verify that $G-h$ is a cograph.  
    Instead of showing that, for
    all $x, y \in X\setminus \{h\}$, it holds that $\lca_T(x,y) = 1$ if and only if $\{x,y\} \in E(G-h)$
	we used    the current proof strategy  which relies solely on the clustering system of $N$.

	Note that Theorem \ref{thm:char-primitive-level1exp} cannot be generalized to primitive
    \levIex[$k$] graphs. As an example, consider the primitive \levIex[$2$] graph $G$ shown in
    Figure~\ref{fig:level2counterex-primitive} where it is necessary to remove at least
    three vertices from $G$ in order to obtain a cograph.

\subsection{Characterizing general {\levIex} graphs}
\label{subS:general}

In this section, we characterize \levIex graphs. As the following result shows, the
property of being a clustering system that is closed or satisfies (L) is hereditary. This fact, in
turn, enables us to prove that the property of a graph being \levIex is also hereditary.

\begin{lemma}[{\cite[L.~2.1]{LSH:24}}]
	\label{lem:hereditary-cluster}
	Let $\mathfrak{C}$ be a clustering system on $X$ with $|X|>1$ and let $x\in X$. Then, $\mathfrak{C}-x$
	is a	clustering system on $X\setminus \{x\}$. If $\mathfrak{C}$ satisfies property $\Pi\in \{closed,
	\text{(L)}\}$, then $\mathfrak{C}-x$ satisfies $\Pi$. 
\end{lemma}

\begin{proposition}\label{prop:lev1ex-hereditary}
	A graph $G$ is \levIex
	if and only if $G[W]$ is \levIex for all non-empty subsets $W\subseteq V(G)$. 
\end{proposition}
\begin{proof}
	The \emph{if} direction follows from the fact that $G[W]$ is \levIex for $W= V(G)$ and that, in this case,  $G[W]= G$.
	Suppose now that $G=(X,E)$ is \levIex and thus, explained by a $0/1$-labeled level-1 network $(N,t)$ on $X$. 
	Let $W\subseteq X$ be non-empty.
	Put $\mathfrak{C}'\coloneqq \mathfrak{C}_N\cap W = \{C\cap W \colon C\in \mathfrak{C}_N \text{ and } C \cap W\neq \emptyset\}$.
	Since $N$ is a level-1 network, Theorem~\ref{thm:CharLvl1Cluster} implies that 
	$\mathfrak{C}_N$ is closed and satisfies (L).	
	Note that for
	$\overline W=\{x_1,\ldots,x_\ell\}= X\setminus W$ it holds that $C\cap
	W=C\setminus \overline W=(\ldots((C\setminus\{x_1\})\setminus\{x_2\})\ldots)\setminus\{x_\ell\}$ for each $C\in\mathfrak{C}_N$. Thus
	$\mathfrak{C}'$ can be obtained as $\mathfrak{C'}=(\ldots((\mathfrak{C}_N-x_1)-x_2)\ldots)-x_\ell$.
	Repeated application of Lemma \ref{lem:hereditary-cluster} to all $x\in X\setminus W$ shows
	that $\mathfrak{C}'$ must be closed and satisfy (L). 
	Theorem~\ref{thm:CharLvl1Cluster} implies that $N'\doteq \Hasse(\mathfrak{C}')$ is a level-1 network on $W$.
	By Lemma \ref{lem:lvl1=>lcaN}, $N'$ is an lca-network.  
	This together with
	Proposition~\ref{prop:lcaN-explain} implies that $N'$ can be equipped with a $0/1$-labeling $t'$ such
	that $(N',t')$ explains $G[W]$. In summary, $G[W]$ is \levIex.	
\end{proof}

	Our goal is to modify the MDT of a \levIex graph $G$ to construct a $0/1$-labeled level-1
    network that explains $G$. To achieve this, we first introduce the concept of ``prime vertex
    replacement'', as originally defined in \cite{BSH:22} and subsequently utilized in
    \cite{HS:22,HS:24,LSH:24}. In this work, we adhere to the terminology and definitions
    established in \cite{LSH:24}.

\begin{definition}[\bf prime-explaining family]\label{def:pfam}
	Let $G = (X,E)$ be a graph with modular decomposition tree  $(\MDT,\tau)$ and 
	$\P\subseteq \MDstrong(G)$ be the set of all of its non-trivial prime modules.
	Two networks $N$ and $N'$ are \emph{internal vertex-disjoint} if
	$(V^0(N)\setminus\{\rho_N\})\cap(V^0(N')\setminus\{\rho_{N'}\})=\emptyset$.
	 A \emph{prime-explaining family (of $G$)} is a set
	$\pfam(G)=\{(N_M,t_M) \colon M\in\P\}$ of pairwise internal vertex-disjoint  $0/1$-labeled networks such
	that, for each $M\in\P$, the network $(N_M,t_M)$ explains the quotient graph $G[M]/\Mmax(G[M])$. 
\end{definition}

As we shall see, the modular decomposition tree $(\MDT,\tau)$ of any graph $G$ can be 
modified by locally replacing prime vertices $M \in \P$ with $(N_M, t_M) \in \pfam(G)$ to construct 
a $0/1$-labeled network that explains $G$. This concept is formalized in the following definition.

\begin{definition}[\bf prime-vertex replacement (pvr) networks]
  \label{def:pvr}
	Let $G=(X, E)$ be a graph and $\P$ be the set of all non-trivial prime vertices in
	its modular decomposition tree $(\MDT,\tau)$. 
	Let $\pfam(G)$ be a prime-explaining family of $G$.
	A \emph{prime-vertex replacement (pvr) network} $\pvr(G,\pfam(G))$ of $G$ is the directed, $0/1$-labeled graph $(N,t)$ obtained by the following procedure: \smallskip
\begin{enumerate}
\item For each $M\in \P$, remove all edges $(M,M')$ with
  		$M'\in \child_{\MDT}(M)$ from $\MDT$ to obtain the directed graph
  		$T'$.\label{step:T'}\smallskip
\item Construct a directed graph $N$ by adding, for each $M\in\P$,  $N_M$ to $T'$ by identifying the root
  		of $N_M$ with $M$ in $T'$ and each leaf $M'$ of $N_M$ with the
	  	corresponding child $M'\in \child_{\MDT}(M)$. \label{step:T''}\smallskip
\item \label{step:color} 
 Define the labeling $t\colon V^0(N)\to \{0,1\}$ by putting, for
  all $w\in V^0(N)$,
  \begin{equation*}
    t(w) \coloneqq
    \begin{cases} 
		 t_{M}(w) &\mbox{if }  w \in V^0(N_{M}) \text{ for some } M\in \P\\
      \tau(w) &\mbox{otherwise, i.e., if } w\in V^0(\MDT)\setminus \P 
    \end{cases}
  \end{equation*}  
\item Replace every singleton module $\{x\}\in V(N)$ in $N$ by $x$.\label{step:flatten leaves}\smallskip
\end{enumerate}
The resulting $0/1$-labeled directed graph  $(N,t)$ is called a \emph{pvr-network (of $G$)} and denoted by $\pvr(G,\pfam(G))$.
\end{definition}

The following proposition establishes that the pvr-network provides a $0/1$-labeled network
explaining any graph $G$, given a prime-explaining family $\pfam(G)$. 
In Lemma~\ref{lem:pfam-lvl-k=>pvr-lvl-k}, which generalizes \cite[L.~6.19]{LSH:24}, 
we further show that pvr-networks preserve certain properties of the networks included in $\pfam(G)$. 
In particular, the level of the pvr-network corresponds to the maximum level of the networks in $\pfam(G)$.

\begin{proposition}[{\cite[L.~5.3 \& Prop.~5.5]{LSH:24}}]\label{prop:pvr-explains-S}
	For every graph $G=(X,E)$, the pvr-network $\pvr(G,\pfam(G))$ is a $0/1$-labeled network on $X$ that explains  
	$G$, for  all  prime-explaining families $\pfam(G)$.
\end{proposition}

\begin{lemma}\label{lem:pfam-lvl-k=>pvr-lvl-k}
	Let $\pfam(G)$ be a prime-explaining family of a graph $G$. 
	If  all $0/1$-labeled networks in $\pfam(G)$ are phylogenetic, then  the pvr-network $\pvr(G,\pfam(G))$ is phylogenetic. 
	If  all $0/1$-labeled networks in $\pfam(G)$ are level-$k$, then  $\pvr(G,\pfam(G))$ is level-$k$.
\end{lemma}
\begin{proof}
	Let $\pfam(G)=\{(N_M,t_M)\colon M\in\P\}$ be a prime-explaining family of the graph $G = (X,E)$, 
	where $\P$ is the set of prime-labeled vertices in the MDT $(\MDT,\tau)$ of $G$.
	Furthermore, put $(N,t)\coloneqq\pvr(G,\pfam(G))$. By Proposition~\ref{prop:pvr-explains-S}, $N$ is a network on $X$. 
	
	Assume first that, for each $(N_M,t_M)\in\pfam(G)$, the network $N_M$ is phylogenetic.
	Since $\MDT$ is phylogenetic and the in- and out-degrees of non-prime vertices in $\MDT$ remain unaffected when
	constructing $N$, it follows that all  vertices	$v$ in $\MDT$ that do not correspond to a prime module of $G$
	do not satisfy $\outdeg_N(v)=1$ and $\indeg_N(v)\leq 1$. Let us now consider a vertex $v$
	in $\MDT$ that corresponds to a prime module $M'$ of $G$. 
	In this case, $v$ is now the root of $N_{M'}$ and thus, $\outdeg_N(v)=\outdeg_{N_{M'}}(\rho_{N_{M'}})>1$ must hold. 	
	Moreover, since $N_{M'}$ is phylogenetic, none of its inner vertices $v$ satisfy $\outdeg_N(v)=1$ and $\indeg_N(v)\leq 1$.
	If, on the other hand, $v$ is a leaf in $N_{M'}$, then it corresponds to some vertex in $\MDT$ which, by the latter arguments, 
	does not satisfy $\outdeg_N(v)=1$ and $\indeg_N(v)\leq 1$. In summary, $N$ is phylogenetic.
	
	Assume now that, for each $(N_M,t_M)\in\pfam(G)$, the network $N_M$ is level-$k$.  
	To verify that $N$ is level-$k$ network,  we must verify that every non-trivial block $B$ of $N$ 
	contains at most $k$ hybrids distinct from $\max_B$.
	Observe that, for any two $M',M\in\P$, the networks $N_M$ and $N_{M'}$ can, by construction, share at most one vertex $v$, 
	namely the root of one of these networks and a leaf of the other network. It follows that, in the latter case, 
	$v$ is a vertex such that $N-v$ is disconnected. This immediately implies that $B$ is a non-trivial block in $N$ if and only if $B$ is a 
	non-trivial block in $N_M$ for precisely one $M\in \P$. This and the fact that all
	non-trivial blocks $B$ in each of the $N_M$ contain at most $k$ hybrids distinct from $\max_B$, implies that every
	non-trivial block $B$ in $N$ contains at most $k$ hybrids distinct from $\max_B$.
	Taking the latter arguments together, $N$ is a level-$k$ network.
\end{proof}

We are, finally, in a position to provide several characterizations of \levIex graphs.

\begin{theorem}\label{thm:general-lev1ex}
  A graph $G$ is {\levIex} if and only if for every non-trivial prime module $M$ of $G$ 
  the quotient graph $G[M]/\Mmax(G[M])$ is a near-cograph. 
\end{theorem}
\begin{proof}
	Let $G$ be \levIex. If $G$ does not contain any non-trivial prime module, 
	then the statement is vacuously true. 
	Thus, suppose that $G$ contains a prime module $M$ with $|M|>1$. 
	Consider the quotient graph $G'\coloneqq G[M]/\Mmax(G[M])$.
	with $\Mmax(G[M]) = \{M_1, \dots , M_k\}$. 
	By Observation~\ref{obs:quotient},
	$G'\simeq G[W]$ with $W\subseteq M$ such that for all $M'\in \Mmax(G[M])$ we have $|M' \cap W | = 1$.
	By Proposition~\ref{prop:lev1ex-hereditary}, $G'$ is \levIex since $G$ is \levIex.
	Since $M$ is a prime module with $|M|>1$, Lemma~\ref{lem:Hprimitive-in-primeM}
	ensures that $G'$ is a primitive graph. The latter two arguments together with Theorem~\ref{thm:char-primitive-level1exp} imply that $G'$ is a near-cograph.

	Now suppose that for every non-trivial prime module $M$ of $G$ 
	the quotient graph $G_M\coloneqq G[M]/\Mmax(G[M])$ is a near-cograph. 
	By Corollary~\ref{cor:near-cograph=>lev1ex}, $G_M$ is \levIex for every non-trivial prime module $M$ of $G$.
	Hence, there exists a prime-explaining family $\pfam$ of $G$ containing only labeled level-1 networks. 
	By Lemma~\ref{lem:pfam-lvl-k=>pvr-lvl-k}, the pvr-network $(N,t)\coloneqq\pvr(G,\pfam)$ is a level-1 network and, 
	by Proposition~\ref{prop:pvr-explains-S}, $(N,t)$ explains $G$. Thus, $G$ is \levIex. 
\end{proof}

\begin{theorem}\label{thm:genaral-lev1ex-v2}
A graph $G$ is \protect{\levIex} if and only if all primitive induced subgraphs of $G$ are near-cographs. 
\end{theorem}
\begin{proof}
	By contraposition, assume that $G$ contains a primitive
	induced subgraph $H=G[W]$ that is not a near-cograph
	and let $M$ be an inclusion-minimal strong module that contains $W$. 
	By Lemma~\ref{lem:Hprimitive-in-primeM}, $M$ is prime
	and $H$ is isomorphic to an induced subgraph of $G[M]/\Mmax(G[M])$.
	By Lemma~\ref{lem:char-near-cograph}, the property of being a near-cograph is hereditary.
	The latter arguments imply that $G[M]/\Mmax(G[M])$
	cannot be a near-cograph. 
	By Theorem~\ref{thm:general-lev1ex},  
	$G$ is not \levIex. 
	This establishes the \emph{only-if} direction. 
	
	Assume now that all primitive induced subgraphs of $G$ (if there are any) 
	are near-cographs. Note that if $G$ does not contain prime modules at all, 
	then Theorem~\ref{thm:general-lev1ex} implies that $G$ is \levIex.
	Hence, assume that $G$
	contains a prime module $M$. By Observation~\ref{obs:quotient}, 
	$G[M]/\Mmax(G[M])\simeq H$ where $H$ is an induced subgraph of $G$. 
	In particular, $H$ is primitive by Lemma~\ref{lem:Hprimitive-in-primeM}.
	By assumption, $H$ is a near-cograph.
	Since the latter argument holds for all prime modules of $G$, 
	Theorem~\ref{thm:general-lev1ex}  implies that $G$ is \levIex. 
	This establishes the \emph{if} direction. 
\end{proof}

Note that primitivity of the induced subgraphs of $G$ cannot be omitted in Theorem~\ref{thm:genaral-lev1ex-v2}.
By way of example, the subgraph $G[W]$ of the  graph $G$ in Figure~\ref{fig:level2counterex} 
that is induced by $W = \{a_1,b_1,c_1,d_1,a_2,b_2,c_2,d_2\}$ consists of two vertex-disjoint $P_4$s
and is, therefore, not a near-cograph even though $G$ is \levIex.

The latter results can now be used to establish
\begin{theorem}\label{thm:lev1ex<=>phylo-lev1-ex}
	A graph is {\levIex} if and only if it can be explained by a \emph{phylogenetic} 0/1-labeled level-1 network.
\end{theorem}
\begin{proof}
	The \emph{if} direction is trivial. Suppose that $G$ is \levIex. If $G$ does not contain any
	non-trivial prime module, then $G$ is a cograph and the cotree that explains $G$ is a
	phylogenetic 0/1-labeled level-1 network. Suppose that $G$ contains prime modules. Let
	$\mathfrak{M}$ denote the set of all non-trivial prime modules $M$ of $G$. By
	Theorem~\ref{thm:general-lev1ex} and Lemma~\ref{lem:Hprimitive-in-primeM}, the quotient graph
	$G_M\coloneqq G[M]/\Mmax(G[M])$ is a primitive near-cograph for all $M\in \mathfrak{M}$. By
	Theorem~\ref{thm:char-primitive-level1exp}, $G_M$ is \levIex for all $M\in \mathfrak{M}$. By
	Lemma~\ref{lem:lvl1+primitive=>unique-hybrid}, there exists a $0/1$-labeled phylogenetic level-1
	network $(N_M,t_M)$ that explains $G_M$ for all $M\in \mathfrak{M}$. Hence, there exists a
	prime-explaining family $\pfam$ of $G$ containing only phylogenetic 0/1-labeled level-1 networks.
	By Lemma~\ref{lem:pfam-lvl-k=>pvr-lvl-k}, the pvr-network $\pvr(G,\pfam(G))$ is a phylogenetic
	0/1-labeled level-1 network. By Proposition~\ref{prop:pvr-explains-S}, $\pvr(G,\pfam(G))$
	explains $G$. This establishes the \emph{only if} direction. 
\end{proof}

\subsection{Linear-time algorithm}
\label{subS:linetimeAlgo}

	We now present an algorithm that determines, in linear time, whether a given graph is \levIex
    and, if so, constructs a $0/1$-labeled level-1 network explaining the input graph.

\begin{algorithm}[tbp]
\small
  \caption{\texttt{Check if $G$ is \levIex and construct Level-1 network that explains $G$}}
\label{alg:general}
\begin{algorithmic}[1]
  \Require Graph $G=(V,E)$
  \Ensure \multiline{%
  					$0/1$-Labeled level-1 network $(N,t)$ that explains $G$, if $G$ is \levIex
  					and,  otherwise, statement ``\emph{$G$ is not \levIex}''
  					}%
	\State  Compute MDT $(\MDT_G,\tau_G)$ of $G$ \label{alg:MDT}
	\State  $\P\gets$ set of $\primeL$-labeled vertices in $(\MDT_G,\tau_G)$ \label{alg:P}
	\If{$\P = \emptyset$} \label{alg:ifCograph}
	 \Return  $(\MDT_G,\tau_G)$
	\EndIf	
	\State $\pfam(G) \gets \emptyset$ \label{alg:init_pfam}
	\State $\mathcal{Q} \gets \{G[M]/\Mmax(G[M]) \colon M\in \P\}$ \label{alg:init_quotients}
	
	\For{all $H\in \mathcal{Q}$} \label{alg:forallH}
		\If{$H$ is not a near-cograph} \label{alg:if_nearCog}
			\State \Return  \emph{$G$ is not \levIex}
		\Else	
		\State Let $z$ be a vertex in $H$ for which $H-z$ is a cograph \label{alg:det_z}
		\State Let $(T^H,t^H)$ be the cotree of $H-z$ \label{alg:cotree}
		\State $\pfam(G) \gets \pfam(G)\cup\{(T^H_{\NWA z},t^H_{\NWA z})\}$ with $(T^H_{\NWA z},t^H_{\NWA z})$ according to Def.~\ref{def:graft}
				and Thm.~\ref{thm:N<-v_explains-G} \label{alg:det_pfam}
		\EndIf	    
	\EndFor
	\State \Return $\pvr(G,\pfam(G))$ \label{alg:return}
\end{algorithmic}
\end{algorithm}
	
\begin{theorem}\label{thm:AlgGeneral}
	It can be verified in $O(|X|+|E|)$ time if a given graph $G=(X,E)$ can
	be explained by a labeled level-1 network and, in the affirmative case, 
	a $0/1$-labeled level-1 network $(N,t)$  that explains $G$  can
	be constructed within the same time complexity. 
\end{theorem}
\begin{proof}	
	To prove the statement, we employ Algorithm~\ref{alg:general} and start with proving its
	correctness.
	Let $G=(X,E)$ denote the input
	graph. The algorithm first computes the modular decomposition tree $(\MDT_G,\tau_G)$ of $G$ and,
	in particular, finds the set $\P$ of $\primeL$-labeled vertices of $\MDT_G$. If
	$\P=\emptyset$, then $\tau_G$ is a $0/1$-labeling of the tree $\MDT_G$ and, therefore, $G$ is a
	cograph and $(\MDT_G,\tau_G)$ is a level-1 network that explains $G$ (c.f.
	Theorem~\ref{thm:CharCograph}). Consequently, $G$ is \levIex and $(\MDT_G,\tau_G)$ is correctly
	returned on Line~\ref{alg:ifCograph}. Henceforth, assume that $\P\neq\emptyset$. Next, an empty
	set $\pfam(G)$ is initialized and the quotient graph $G[M]/\Mmax(G[M])$ is constructed for each
	$M\in \P$, where $\mathcal{Q}$ comprise all such quotients. Due to
	Theorem~\ref{thm:general-lev1ex}, $G$ is \levIex if and only if $H$ is a near-cograph for every
	$H\in\mathcal{Q}$. Therefore, the \emph{for}-loop starting on Line~\ref{alg:forallH} and the
	\emph{if}-clause on Line~\ref{alg:if_nearCog} will correctly determine whenever $G$ is not
	\levIex and thereafter terminate. If, instead, $G$ is \levIex then
	Lines~\ref{alg:det_z}--\ref{alg:det_pfam} will run for each $H\in\mathcal{Q}$. More specifically,
	the cotree $(T^H,t^H)$ of the cograph $H-z$ for some $z\in V(H)$ can be found and it, by definition,
	explains $H-z$. By Lemma~\ref{lem:graft-properties} and Theorem~\ref{thm:N<-v_explains-G},
	there is a $0/1$-labeling $t^H_{\NWA z}$ such that $(T^H_{\NWA z},t^H_{\NWA z})$ is a level-1 network that explains
	$H$. Consequently, after the last iteration of the loop, the set $\pfam(G)$ will be a
	prime-explaining family of $G$ containing only level-1 networks. By
	Proposition~\ref{prop:pvr-explains-S} and Lemma~\ref{lem:pfam-lvl-k=>pvr-lvl-k},
	$\pvr(G,\pfam(G))$ is therefore a level-1 network that explains $G$. This network is, correctly,
	returned in Line~\ref{alg:return}.

	We now consider the runtime.
	Computation of the MDT $(\MDT_G,\tau_G)$ of $G$ in Line~\ref{alg:MDT} can be achieved in $O(|X|+|E|)$
	time \cite{Habib:2010}. The number of vertices in $\MDT_G$ is in $O(|X|)$ \cite{Ehrenfeucht:1994} and thus,
	the set $\P$ of prime vertices in $(\MDT_G,\tau_G)$ can be determined in $O(|X|)$ time
	(Line~\ref{alg:P}). The tasks in Line~\ref{alg:ifCograph} and \ref{alg:init_pfam} take constant
	time. As shown in \cite[L.~2]{DUCOFFE2021201}, \emph{all} quotients $G[M]/\Mmax(G[M])$ and thus,
	the set $\mathcal{Q}$ can be computed $O (|X| + |E|)$ time. Note that $\mathcal{Q}$ has at most
	$|V(\MDT_G)|$ elements. 

	We consider now the runtime of one iteration of the for-loop starting in Line \ref{alg:forallH}. 
	To better distinguish between strong modules in  $\MDstrong(G)$  and vertices in $(\MDT_G,\tau_G)$ (which are the elements in  $\MDstrong(G)$)
		we use another notation. Let  $H\in \mathcal{Q}$ as in Line~\ref{alg:forallH}. 
		Then  $H =	G[M]/\Mmax(G[M])$ for some module 	$M\in \MDstrong(G)$ which is also vertex in  $(\MDT_G,\tau_G)$
		and we denote this vertex in the following by $v_H$. Moreover, put $n_H \coloneqq
	|\child_{\MDT_G}(v_H)|$ and note that $H$ contains exactly $n_H$ vertices, since $V(H)=\Mmax(G[M])=\child_{\MDT_G}(v_H)$. Let $m_H$ denote the
	number of edges in $H$. We first argue that determining if $H$ is a near-cograph and, in the
	affirmative case, to find a vertex $z$ resulting in the cograph $H-z$ can done in $O(n_H+m_H)$
	time (Line~\ref{alg:if_nearCog} - \ref{alg:det_z}). To achieve this goal, we can use the $O(n_H +
	m_H)$-time algorithm of Corneil et al.\ \cite{Corneil:85} that is designed to verify if a given
	graph is a cograph or not. Here, we use $H$ as input for this algorithm. The algorithm is
	incremental and constructs the cotree of a subgraph $H[V']$ of $H$ starting from a single vertex
	and then increasing $V'$ by one vertex at each step of the algorithm. Since $H$ is not a cograph,
	at some point there is a set $V'$ and a chosen vertex $w$ such that $H[V']$ is a cograph and
	$H[V'\cup \{w\}]$ contains an induced $P_4$. Based on these findings, Capelle et al.\
	\cite{capelle1994cograph} showed that one can find an induced $P_4$ containing $w$ in
	$O(\deg_H(w)) \subseteq O(n_H)$ time. Henceforth, we denote such an induced $P_4$ just by $P$.
	Now, we aim at finding a vertex $z$ such that $H-z$ is a cograph. Due to Observation~\ref{obs:allP4s}, 
	$z$ must be located on
	$P$, i.e., it is one of the four vertices in $P$. Thus, we check for all four vertices $z$ in $P$
	again if $H-z$ is a cograph with the $O(n_H + m_H)$-time algorithm of Corneil et al. In summary,
	the tasks in Line~\ref{alg:if_nearCog} - \ref{alg:det_z} can be achieved in $O(n_H + m_H)$ time.
	The worst case for the runtime is of course the case that all graphs in $\mathcal{Q}$ are
	near-cographs. Thus, suppose that $H-z$ is a cograph. In this case, we can assume that the cotree
	$(T^H,t^H)$ of $H-z$ as required in Line~\ref{alg:cotree} is already computed in the previous
	step. We now compute $(T^H_{\NWA z},t^H_{\NWA z})$ and add it to $\pfam(G)$ in
	Line~\ref{alg:det_pfam}. Note that $T$ has precisely the vertices of $H$ as its leaves and it is
	an easy task to verify that, therefore, $T^H_{\NWA z}$ can be constructed in $O(n_H)$ time.
	Keeping the labels of all $O(n_H)$ vertices in $T^H$ that are present in $T^H_{\NWA z}$ and
	adding labels to the $O(n_H)$ newly created vertices allows us to construct $t^H_{\NWA z}$ in
	$O(n_H)$ time. In summary, the time complexity to compute all tasks within a single iteration of the for-loop is
	in $O(n_H + m_H)$ for each $H\in \mathcal{Q}$. 
   
   We consider now the overall runtime of the for-loop. By construction, $|\mathcal{Q}| = |\P| \leq |V(\MDT_G)|$.
   Hence, the the overall runtime of the for-loop can be expressed as $O(\sum_{v_H\in \P} (n_H +
   m_H)) = O(\sum_{v_H\in \P} n_H) + O(\sum_{v_H\in \P} m_H)$. As argued above, $n_H =
   |\child_{\MDT_G}(v_H)|$ and hence, $O(\sum_{v_H\in \P} n_H )= O(\sum_{v_H\in \P}
   |\child_{\MDT_G}(v_H)|) \subseteq O(|E(\MDT_G)|) = O(|V(\MDT_G)|) = O(|X|)$. 
   In the following, we argue that  $O(\sum_{v_H\in \P} m_H) \subseteq O(|E|)$.
   To this end, let $\{M_i,M_j\}$ be an edge in $H$ for some $v_H\in \P$. 
   This edge exists in $H$ because there is an edge $\{x,y\}\in E$ with $x\in M_i$ and $y\in M_j$. 
   Let us call such an edge $\{x,y\}$ a witness for the edge $\{M_i,M_j\}$.
   Hence, any edge in $H$ is witnessed by an edge in $E$, i.e., $m_H\leq |E|$ for all $v_H\in P$. 
   Moreover, since any two distinct prime modules $M$ and $M'$ of $G$ are strong, 
   it holds that (i) $M\cap M'  = \emptyset$ or (ii) $M\subsetneq M'$ or (iii)  $M'\subsetneq M$.
   In case (i) and (ii) we have $x\notin M'$ or $y\notin M'$. In case (ii), 
   we have $\{x,y\}\subseteq M_k$ for some inclusion-maximal strong module $M_k\subsetneq M'$. 
   In either case, the existence of any edge $\{M'_i,M'_j\}\in E(H')$ with $H' = G[M']/\Mmax(G[M'])$ 
   is not witnessed by any edge that witnessed an edge in $H$. 
   In other words, the set of edges that witness an edge in $H$ and 
   the set of edges that witness an edge in $H'$ are disjoint. 
	Thus, $O(\sum_{v_H\in \P} m_H) \subseteq O(|E|)$.  In summary, the overall runtime of the for-loop is
	in $O(|X|+|E|)$. 
   
   By construction, the network $(T^H_{\NWA z},t^H_{\NWA z})$ for every $H\in \mathcal{Q}$ contains
   the vertices of the MDT $\MDT_H$ of $H$ whose leave set is of size $n_H = |\child_{\MDT_G}(v_H)|$
   and $O(n_H)$ additional vertices. It is now easy to verify that modifying $(\MDT_G,\tau_G)$ to
   $\pvr(G,\pfam(G))$ can be done in $\sum_{v_H\in \P} n_H = \sum_{v_H\in \P} |\child_{\MDT_G}(v_H)|
   \in O(|X|)$ time. 
   
   In summary, can be verified in $O(|X|+|E|)$ time if a given graph $G=(X,E)$ can be explained by a
   labeled level-1 network and, in the affirmative case, a $0/1$-labeled level-1 network $(N,t)$
   that explains $G$ can be constructed within the same time complexity.
\end{proof}

\subsection{The substitution operator, perfect graphs, twin-width and related results}

The latter results allow us to show that every \levIex graph is obtained from 
\levIex graph by ``locally replacing'' vertices by \levIex graphs. To this end, we need the following

\begin{definition}
For two vertex-disjoint graphs $G$ and $H$ and a vertex $v\in V(G)$ we define 
the \emph{substitution operation $G \inflateOP_v H$} as the graph defined by
\begin{enumerate}[noitemsep, label=\arabic*.]
	\item Put $G' \coloneqq (G-v)\union H$ and
	\item add the edge $\{u,w\}$ to $E(G')$ for all $\{u,v\}\in E(G)$ and for all $w\in V(H)$
	      which results in  $G \inflateOP_v H$.
\end{enumerate}
	In this case, we say that $v$ \emph{is substituted by $H$ (in $G$)}. 
\end{definition}

In simple terms, $G \inflateOP_v H$ replaces the vertex $v$ in $G$ with the graph $H$, where the
vertices in $H$ are adjacent exactly to those vertices in $G$ to which $v$ was originally adjacent.
The substitution operator is a standard tool in graph theory and has been heavily used, in
particular, to prove the ``perfect graph theorem''  \cite{Lovasz:1972, LOVASZ:72b} and to
establish results related to primitive graphs or hereditary graph classes, see,
e.g., \cite{DRGASBURCHARDT:11,GIAKOUMAKIS:07,BRIGNALL201960,ZVEROVICH:03}.
In particular,
Drgas-Burchardt \cite{DRGASBURCHARDT:11} provides a characterization of those hereditary graph
classes that are closed under the substitution operator. These results depend on some notions
that differ significantly from our notation. Hence, instead of reusing these results, we provide
a direct proof to ensure self-consistency.

\begin{theorem}\label{thm:char-LevIex-inflate}
	The class of {\levIex} is closed under the substitution-operation $\inflateOP$. 
	In particular,
	a graph $G$ is \levIex if and only if 
	\begin{enumerate}[noitemsep, label=(\roman*)]
	\item $G\simeq K_1$\  or\  (ii) $G = H \inflateOP_v H'$ with $H$ and $H'$ being \levIex and $|V(H')|>1$.
	\end{enumerate}
	Furthermore, the disjoint union and the join of two \levIex graph is a \levIex graph.
\end{theorem}
\begin{proof}
	Suppose that $G=(X,E)$ is \levIex. If $|X|=1$ then $G\simeq K_1$ satisfies (i). Assume that
	$|X|>1$. In this case, let $K_1 = (\{v\},\emptyset)$. Thus, 
	$G = K_1 \inflateOP_v G$ and 
	$G$ satisfies (ii).
	Suppose that $G = (X,E)$ satisfies (i) or (ii). If $G \simeq K_1$, then it is clearly
	\levIex. Suppose that $G = H \inflateOP_v H'$
	for two \levIex graphs $H$ and $H'$, and a vertex $v\in V(H)$.
	If $G$ contains no prime module, then Theorem~\ref{thm:general-lev1ex} implies that $G$ is \levIex.
	Assume now that $M$ is a prime module of $G$ and put $W = V(H')$. 
	By Lemma~\ref{lem:Hprimitive-in-primeM}, $M$ is strong. 
	Note that all vertices in $W$
	have in $G$, by construction, the same neighbors and non-neighbors  in $X\setminus W$, namely
	those of $v$ in $H$. Hence, $W$ is a module in $G$. Since $M$ is strong, it does not overlap with
	$W$. Hence, there are three cases that we consider $M\cap W = \emptyset$, $W\subsetneq M$ or
	$M\subseteq W$. If $M\cap W = \emptyset$, then $G[M] = H[M]$ and thus,
	$\Mmax(G[M]) = \Mmax(H[M])$. Hence, $G[M]/\Mmax(G[M]) = H[M]/\Mmax(H[M])$. Since $H$ is \levIex,
	Proposition~\ref{prop:lev1ex-hereditary} implies that $H[M]$ is \levIex. In particular, 
	$M$ remains a prime module in $H[M]$. Thus,
	Theorem~\ref{thm:general-lev1ex} implies that $H[M]/\Mmax(H[M]) = G[M]/\Mmax(G[M])$ is a near-cograph. 
	If $M\subseteq W$, then $G[M] = H'[M]$ and we can reuse the latter arguments applied on $H'$
	instead of $H$ to conclude that $G[M]/\Mmax(G[M])$ is a near-cograph. 
	If $W\subsetneq M$ then we consider $\Mmax(G[M]) = \{M_1,\dots,M_k\}$. Since all $M_i\in \Mmax(G[M])$
	are strong, they do not overlap with $W$. This and  $W\subsetneq M$ implies that 
	that $W\subseteq M_i$ for some $M_i\in \Mmax(G[M])$
	which, in particular, implies that $H[M_j] = G[M_j]$ for all $j\in \{1,\dots k\}\setminus \{i\}$, 
	and that the adjacencies between vertices in $M_j$ and $M_l$ remain unchanged for all $j,l \in \{1,\dots k\}\setminus \{i\}$.
	Moreover, by construction,	$X = (V(H)\setminus \{v\} )\cup W$ and
	the adjacent vertices of  $w \in W$ in $G[X\setminus W]$ 
	are exactly the adjacent vertices of  $v$ in $H[V(H)\setminus \{v\}]$. 
	The latter arguments imply that
	\[\Mmax(H[M])=\{M_1,\ldots, M_{i-1},(M_i\setminus W)\cup\{v\},\ldots,M_k\}\]
	and  $G[M]/\Mmax(G[M]) \simeq H[M]/\Mmax(H[M])$ which, by Theorem~\ref{thm:general-lev1ex}, implies that $G[M]/\Mmax(G[M])$ is a near-cograph.
	Hence, for each of the cases  $M\cap W = \emptyset$, $W\subsetneq M$ and
	$M\subseteq W$, the graph
	$G[M]/\Mmax(G[M])$ is a near-cograph and this holds for every prime module
	$M$ of $G$. Theorem~\ref{thm:general-lev1ex} implies that $G$
	is \levIex. The latter arguments imply that the class of \levIex is closed under the substitution-operation $\inflateOP$.
   
   Finally, assume that  $G$ is the join or disjoint union of two \levIex graphs  $G_1$ and $G_2$. 
   Suppose that $G$ contains a prime module $M$. Hence, neither $G[M]$
   nor $\overline{G}[M]$ is connected. This directly implies that $M\subseteq V(G_i)$
   for one $i\in \{1,2\}$. Therefore, $G[M] = G_i[M]$ and thus, $G[M]/\Mmax(G[M]) = 
   G_i[M]/\Mmax(G_i[M])$. By Theorem~\ref{thm:general-lev1ex}, $G[M]/\Mmax(G[M])$ a near-cograph. 
   As the latter arguments hold for all prime modules $M$ of $G$, 
   Theorem~\ref{thm:general-lev1ex} implies that $G$ is \levIex.
\end{proof}

We consider now `atomic expressions'' $\atomE$ of $G$ that are based on the operations $\inflateOP$, $\union$, and $\join$. 
As already shown in Theorem~\ref{thm:char-LevIex-inflate}, every \levIex graph $G$ can be written as 
a sequence of \levIex graphs in which all graphs are concatenated by the $\inflateOP$-operator. 
We show now that every \levIex graph can be written as a sequence of $K_1$'s and primitive near-cographs
where the graphs are concatenated by the $\inflateOP$-, $\union$- and $\join$-operator.
To be more precise:

\begin{definition}
    For a graph class $\mathcal{G}$, we write $\underline{\mathcal{G}}$ to denote that one graph from $\mathcal{G}$ is taken.  
    We denote by $\mathcal{N}$ the set of all primitive near-cographs and by $\mathcal{K}$ the set of all single-vertex graphs $K_1$. 
    Moreover, when using $H \inflateOP_v H'$, we always assume that $v$ is a vertex in $H$.
    In addition, when using $H \inflateOP_v H'$, $H \union H'$, or $H \join H'$, we always assume that $H$ and $H'$ are vertex-disjoint.
		
	\noindent	
    An \emph{atomic expression} ($\atomE$) is an expression defined recursively by the following grammar:
	\begin{equation} \label{eq:atomE}
		\begin{split}
		\atomE &\rightarrow \mathsf{A} \mid ( \atomE \, \join \, \atomE ) \mid ( \atomE \, \union \, \atomE ) \mid ( \atomE \, \inflateOP_v \, \atomE) \\
		\mathsf{A} &\rightarrow  \underline{\mathcal{N}}\, \mid \,\underline{\mathcal{K}}
		\end{split}
	\end{equation}
	\noindent
	An atomic expression $\atomE$ is \emph{valid for $G$} if, after substituting the elements  $\underline{\mathcal{K}}$ and $\underline{\mathcal{N}}$ at each step with 
	specified graphs from the respective classes $\mathcal{K}$ and $\mathcal{N}$,
	the result is a graph $H$ that is isomorphic to $G$. 
	In this case, we also say that $\atomE$ is a \emph{valid} atomic expression for $G$. 
	Moreover, we write $\atomE(G)$ for one of the valid atomic expression for $G$, if it exists.
\end{definition}

We provide now a couple of examples for graphs that have a valid atomic expression. 
The simplest one is provided by the single vertex graph $K_1$ for which both
$\underline{\mathcal{K}}$ and $(\underline{\mathcal{K}} \inflateOP_v \underline{\mathcal{K}})$
are valid atomic expressions, illustrating that there can be multiple valid atomic expressions for the same graph.

Now, consider the graph $G$ as in Figure~\ref{fig:blocks-modules} and let $P = G[\{a,b,c,d\}] \in \mathcal{N}$. 
Moreover, put $K^v \coloneqq (\{v\},\emptyset)$ to specify the vertex set of a $K_1\in \mathcal{K}$. 
In this example, \[G \simeq  (\underline{\mathcal{K}} \join  (\underline{\mathcal{K}}\union(\underline{\mathcal{K}} \join (\underline{\mathcal{K}}\union \underline{\mathcal{K}}))) \inflateOP_v \underline{\mathcal{N}} )  \] 
where the specified graph $\underline{\mathcal{N}}$ is an induced $P_4$ and, in particular,
\[\atomE(G) =  (K^h\join   ((K^v\union(K^x \join (K^y\union K^z))) \inflateOP_v P ))  \] is a valid atomic expression of $G$.
In other words, $G$ is composed of $K_1$s and primitive near-cographs linked by $\join, \union$  and  $\inflateOP$.
Another valid atomic expressions for $G$ is $(((K^v \inflateOP_v P)\union(K^x \join (K^y\union K^z)))\join K^h)$.

The graph $G$ in Figure~\ref{fig:level2counterex} is just the disjoint union of primitive near-cographs, 
namely $P_4$s and we obtain the valid atomic expression \[\atomE(G) = \underbrace{(\dots ((\underline{\mathcal{N}} \union \underline{\mathcal{N}}) \union \underline{\mathcal{N}}) \dots \union \underline{\mathcal{N}})}_{k \text{ times} },\]
where each specified graph $\underline{\mathcal{N}}$ is an induced $P_4$. 

Clearly, every graph can be written as $G = K \inflateOP_v G$ with $K = (\{v\}, \emptyset)$.  
However, not all graphs admit a valid atomic expression. For example, an induced cycle $C_5$ on five vertices  
is primitive but not \levIex (cf.\ Lemma~\ref{lem:odd-hole-free}). Hence, $C_5 \notin \mathcal{K} \cup \mathcal{N}$.  
In particular, it is easy to verify that an induced $C_5$ has no valid atomic expression.  
In contrast, an induced cycle $C_4$ has a valid atomic expression, namely  
$((\underline{\mathcal{K}} \union \underline{\mathcal{K}}) \join (\underline{\mathcal{K}} \union \underline{\mathcal{K}}))$.

\begin{theorem}\label{thm:atomic-expression}
	A graph $G$ is {\levIex} if and only if there exists a valid atomic expression $\atomE(G)$ for $G$. 
\end{theorem}
\begin{proof}
	We prove the \emph{only-if} direction by induction on the number of vertices in $G = (X,E)$. 
	If $|X|=1$, then $\atomE(G) = \underline{\mathcal{K}}$ is a valid atomic expression of $G$. 
	Suppose now that all \levIex graphs $G$ with $|X|\leq n$ vertices have a valid atomic expression $\atomE(G)$ 
	for some $n\geq 1$.
	Let $G = (X,E)$ be a graph with $|X| = n+1 > 1$ vertices and consider the root $\rho$ of the MDT $(\MDT,\tau)$ of $G$. 
	Since $|X|>n$, it follows that for the set of children $\Mmax(G) = \{M_1,\dots,M_k\}$ of $\rho$ it holds
	that $k>1$ and $|M_i|\leq n$. By Proposition~\ref{prop:lev1ex-hereditary}, 
	$G_i\coloneqq G[M_i]$ is \levIex for all $i\in \{1,\dots,k\}$. 
	By induction hypothesis,  $G_i$ has a valid atomic expression $\atomE(G_i)$ for all $i\in \{1,\dots,k\}$. 

	We consider now the three possible labels of $\rho$. If $\tau(\rho) = 0$ then $G = G_1\union
	G_2\union G_3 \ldots \union G_k$ (cf.\ Observation~\ref{obs:parallel-series}) and it follows
	that $\atomE(G) = (\dots((\atomE(G_1)\union \atomE(G_2))\union \atomE(G_3))\dots \union
	\atomE(G_k))$ is a valid expression for $G$. If $\tau(\rho) = 1$ then $G = G_1\join G_2\join G_3
	\ldots \join G_k$ (cf.\ Observation~\ref{obs:parallel-series}) and it follows that
	$\atomE(G) = (\dots((\atomE(G_1)\join \atomE(G_2))\join \atomE(G_3))\dots \join \atomE(G_k))$ is
	a valid expression for $G$. Assume now that $\tau(\rho) = \primeL$. Since $G$ is \levIex,
	Lemma~\ref{lem:Hprimitive-in-primeM} implies that $H\coloneqq G[X]/\Mmax(G[X])= G/\Mmax(G)$ is a
	primitive \levIex graph and, by Theorem~\ref{thm:char-primitive-level1exp}, a primitive
	near-cograph. Consider first the possibility that $|V(H)|=n+1$. In this case,
	$|V(H)|=|\Mmax(G)|=|X|$ which enforces $|M_i|=1$ for each $M_i\in\Mmax(G)$ and, therefore,
	$G\simeq H$. In particular, $\underline{\mathcal{K}}\inflateOP_v\underline{\mathcal{N}}$ is a
	valid atomic expression for $G$, obtained by specifying $\underline{\mathcal{N}}$ to be the
	primitive near-cograph $H$ and $\underline{\mathcal{K}}$ to be the $K_1$ defined as
	$(\{v\},\emptyset)$. Henceforth assume that $|V(H)|\leq n$. By induction hypothesis, 
	there is a valid atomic expression $\atomE(H)$ of $H$. Moreover
	$G$ is obtained from $H$ by
	substituting each vertex $M_1,\dots,M_k$ by $G_1,\dots,G_k$,
	respectively. In other words, $G = ( \dots ((H\inflateOP_{M_1} G_1) \inflateOP_{M_2} G_2) \dots)
	\inflateOP_{M_k} G_k$. Hence, $\atomE(G) = (( \dots ((\atomE(H)\inflateOP_{M_1} \atomE(G_1))
	\inflateOP_{M_2} \atomE(G_2)) \dots) \inflateOP_{M_k} \atomE(G_k))$ is a valid atomic expression
	of $G$. 

	We now proceed with the \emph{if} direction by induction on the number of terminal symbols in
	$\atomE(G)$, that is, on the number of occurrences of $\underline{\mathcal{K}}$ and
	$\underline{\mathcal{N}}$ in $\atomE(G)$. In what follows, we make frequent use of the fact that,
	for every atomic expression $\atomE$, there is a graph $G$ with $\atomE(G) = \atomE$. If
	$\atomE(G)=\underline{\mathcal{K}}$ then $G$ is isomorphic to a $K_1$ and thus clearly \levIex.
	If $\atomE(G)=\underline{\mathcal{N}}$, then $G$ is a primitive near-cograph and by
	Theorem~\ref{thm:char-primitive-level1exp} also \levIex. The latter two arguments comprise the
	base case of one terminal symbol appearing in $\atomE(G)$. Assume that, for some $n\geq 1$
	and for every atomic expression $\atomE(G)$ with at most $n$ terminal symbols, the graph $G$ is a
	\levIex graph. Now let $G$ be a graph with a valid atomic expression $\atomE(G)$ that contains $n+1$
	terminal symbols. By definition of atomic expressions and since $n+1>1$, we have $\atomE(G)=(\atomE'\join\atomE'')$,
	$\atomE(G)=(\atomE'\union\atomE'')$ or $\atomE(G)=(\atomE'\inflateOP_v\atomE'')$ for some atomic
	expressions $\atomE'$ and $\atomE''$. Let $G'$ and $G''$ be two graphs for which $\atomE(G')
	= \atomE'$ and $\atomE(G'') = \atomE''$. Since $\atomE'$
	and $\atomE''$ have at least one terminal symbol each, neither have more than $n$ terminal
	symbols and we can apply the induction hypothesis to conclude that $G'$ and $G''$ are \levIex
	graphs. Hence it follows that $G\simeq G'\join G''$, $G \simeq G'\union G''$ or $G \simeq G'\inflateOP_v G''$. In all
	three cases, Theorem~\ref{thm:char-LevIex-inflate} implies that $G$ is \levIex. 
\end{proof}

We note in passing that the disjoint union and join of two graphs $G$ and $G'$  can alternatively be written
	 as $G\union G'=((\overline{K_2}\inflateOP_v G)\inflateOP_u G')$ and $G\join G'=((K_2\inflateOP_v G)\inflateOP_u G')$
	 where $K_2 = (\{u,v\}, (\{u,v\}))$ is the graph with two vertices $u$ and $v$ connected by an edge.
	 Let us denote with $\mathcal{K}^+$ the set of all graphs isomorphic to $K_1$, $K_2$ and $\overline{K_2}$.
	 Using the notation introduced in Equation~\ref{eq:atomE}, we can thus write both
	 $\atomE \, \join \, \atomE$
	 and $\atomE \, \union \, \atomE$
	 as $(\underline{\mathcal{K}^+} \inflateOP_v \atomE) \,\inflateOP_v \atomE$. 
	 This, together with the derivation ``$\atomE \rightarrow \mathsf{A} \rightarrow \underline{\mathcal{K}^+}$'' and Theorem~\ref{thm:atomic-expression}, 
	 leads to the following result, which essentially mirrors \cite[Cor.~1]{DRGASBURCHARDT:11} for \levIex graphs.

\begin{corollary}
	A graph $G$ is {\levIex} if and only if there exists an valid expression for $G$ in the grammar recursively defined by
	\begin{align*}
		\atomE &\rightarrow \mathsf{A} \mid ( \atomE \, \inflateOP_v \, \atomE) \\
		\mathsf{A} &\rightarrow  \underline{\mathcal{N}}\, \mid \,\underline{\mathcal{K}^+}.
	\end{align*}
\end{corollary}

We close this section by showing the relationship between \levIex graphs and other graph classes, as well as their connection 
to the ``twin-width''.
A graph $G$ is \emph{perfect} if the chromatic number of every induced subgraph equals the size of
the largest clique in that subgraph. It is well-known that perfect graphs are closed under the
substitution operation (cf.\ \cite[Thm~1]{Lovasz:1972}). By Theorem~\ref{thm:char-LevIex-inflate},
\levIex graphs also possess this property. Hence, it is natural to ask whether \levIex graphs are
perfect. 
An affirmative answer to this question is provided in Theorem~\ref{thm:levIex=>perfect}. To
establish this result, we first define a \emph{hole} as an induced cycle $C_n$ on $n \geq 5$
vertices. The complement of a hole is called an \emph{anti-hole}.

\begin{lemma}\label{lem:odd-hole-free}
	\levIex graphs are hole-free and anti-hole-free.
\end{lemma}
\begin{proof}
	By contraposition, suppose that $G$ contains an induced $C_n$ on $n\geq 5$ vertices,
	i.e.,  $G[W]\simeq C_n$ for some $W\subseteq V(G)$. 
	One easily verifies that $G[W]$ is primitive
	and not a near-cograph.
	By Lemma~\ref{thm:char-primitive-level1exp}, 
	 $G[W]$ is not \levIex and, by Proposition~\ref{prop:lev1ex-hereditary},
	 $G$ is not \levIex. Hence, \levIex graphs are hole-free.
	 Assume now that $G$ is \levIex. 
	 By Observation~\ref{obs:complement}, 
	$\overline G$ is \levIex. By the latter arguments, 
	$\overline G$ must be hole-free which 
	implies that $G$ is anti-hole-free.
\end{proof}

A graph is \emph{weakly-chordal} if and only if it does not contain holes or
anti-holes \cite{hayward1985weakly}.
As shown by Hayward in \cite{hayward1985weakly}, weakly-chordal graphs
are perfect. The latter together with Lemma~\ref{lem:odd-hole-free} implies

\begin{theorem}\label{thm:levIex=>perfect}
	\levIex graphs are weakly-chordal and thus, perfect. 
\end{theorem}

Bonnet et al.\ recently introduced a novel parameter called ``twin-width ($\tww$)'' as a
non-negative integer measuring a graphs distance to being a cograph
\cite{Bonnet:2021A} that is based one the following characterization of cographs: A
graph is a cograph if it contains two vertices with the same neighborhood
(called twins), identify them, and iterate this process until one ends in a
$K_1$ \cite{Bonnet:2021A,Bonnet:2024}. This makes cographs the unique class of graphs having twin-width $0$. 
We omit the (lengthy) formal definition of twin-width here and instead refer the reader to \cite{Bonnet:2021A}.

\begin{proposition}
	If $G$ is a \levIex graph, then $\tww(G)\leq 2$. There exists \levIex graphs with twin-width equal to $2$.
\end{proposition}
\begin{proof}
	We first remark that \cite[Thm~4.1]{Bonnet:2021A} states that $\tww(G)\leq 2(\tww(G-v)+1)$ for all graphs $G$ and all vertices $v$ of $G$.
	Since every near-cograph $H$ has, by definition, a vertex $v$ such that $H-v$ is a cograph and thus such that $\tww(H-v)=0$, 
	every near-cograph $H$ satisfies $\tww(H)\leq 2$.
	Furthermore note that, by \cite[Thm.~3.1]{Schidler:2022}, we have 
	\[\tww(G) = \max \{\tww(G[M]/\Mmax(G[M]))\,:\,M \text{ is a prime module of $G$}\},\]
	for all graphs $G$ and thus Theorem~\ref{thm:general-lev1ex} implies that the twin-width of any \levIex graph $G$ equals 
	the maximum twin-width of a set of near-cographs. As previously argued, each such near-cograph has twin-width at most 2, thus so does $G$.
	In summary,  $\tww(G)\leq 2$ for all \levIex graphs $G$.

	We show now that there are \levIex graphs for which this bound is tight. To this end,
	consider the graph $G=(V,E)$ where $V=\{x,v_1,v_2,v_3,u_1,u_2,u_3\}$ and whose edges are
	$\{x,v_i\}$ and $\{v_i,u_i\}$ for $i=1,2,3$. Since $G-x$ is a cograph, $G$ is a near-cograph
	and Corollary~\ref{cor:near-cograph=>lev1ex} implies that $G$
	is \levIex. We show now that $\tww(G)=2$. The  graph $G$ contains an \emph{asteroidal triple} $(u_1,u_2,u_3)$, i.e., 
	$u_1,u_2,u_3$ are pairwise non-adjacent vertices and 
	$u_i$ and $u_j$ are connected by a path in $G$ which does not contain vertices that are adjacent to the vertex $u_k$, $\{i,j,k\} = \{1,2,3\}$. 
	\cite{GOLUMBIC1984157}. In particular, Theorem~4 in \cite{GOLUMBIC1984157} implies that 
	$G$ is not a so-called co-comparability graph. Therefore, \cite{Dushnik:1941} (c.f. \cite[Thm.~4.7.1]{Brandstadt:1999})
	implies that $G$ is not a so-called permutation graph.
	Now Theorem~8 in \cite{Ahn:2025} implies that
	$\tww(G)\geq2$. Since $G$ is \levIex the previous arguments imply that
	$\tww(G)=2$.
\end{proof}

\section{Summary and outlook}
\label{sec:outro}

In this contribution, we have established a general framework for evaluating the biological
feasibility of orthology graphs. In particular, we examined whether a network-based evolutionary
scenario exists that allows for speciation events, represented by a $0/1$-labeling of its inner
vertices, that give rise to the orthologs encoded in the orthology graph. We demonstrated that for
every orthology graph $G$, there exists a network $N$ with a $0/1$-labeling that explains $G$, where
$N$ has no more hybrid vertices than $G$ has vertices (cf. Theorem~\ref{thm:every_G_lev-k-ex}). In
this sense, \emph{every} inferred set of orthologous gene pairs can, in principle, be represented
within some network-based evolutionary scenario. However, such a scenario may be excessively complex
and biologically implausible. It is therefore natural to impose constraints on the complexity of the
permissible networks. To this end, we focused on level-1 networks, which, in an intuitive sense, are
close to being tree-like while still allowing for occasional non-tree-like evolutionary events. We
provided several characterizations of \levIex graphs, i.e., orthology graphs that can be explained
by level-1 networks. In this context, modular decomposition proved to be a useful tool for exploring
the non-tree-like behavior of networks explaining a graph $G$, as such behavior is determined by
the structural properties within
the prime modules of $G$. In particular, $G$ is \levIex if and only if each primitive subgraph
is a near-cograph. Clearly, if 
an orthology graph $G$ lacks this structure, it would be necessary to determine the appropriate
edits (removal or addition of edges) required to transform it into a \levIex graph. 
The computational complexity of the underlying  editing problem remain an interesting open question for future research. 

We propose that modular decomposition can serve as a toolbox for further generalizations,
particularly for characterizing orthology graphs that can be explained by network types beyond
level-1. Specifically, if the primitive subgraphs of an orthology graph $G$ can be explained by a
given network type, then it is plausible that the entire graph $G$ can also be explained by the same
type of network. A reasonable approach to proving this assumption for certain network types is to
employ prime-vertex replacement networks (cf. Definition~\ref{def:pvr}). These networks are used to
replace prime vertices in the modular decomposition tree of $G$ with specified networks that explain
the underlying primitive subgraphs. However, even the natural generalization to level-2 networks
appears to be non-trivial. As illustrated in Figure~\ref{fig:level2counterex-primitive}, additional
hybridization events introduce significantly greater structural complexity to the underlying
orthology graph. Nevertheless, understanding these generalizations could provide deeper insights
into how hybridization affects the representation of orthology relationships.

Another important aspect concerns the relationship between orthologs and best matches. This topic
has received considerable attention in recent years under the assumption of tree-like evolution
\cite{Geiss:19a,BMG-corrigendum,Geiss2020-recons}. An orthology graph that can be explained by a
tree must always be a subgraph of the (reciprocal) best match graph underlying this tree \cite{Geiss2020-recons}. Moreover,
in \cite{Schaller:20x} it was shown that, given the correct best match graph, it is possible to
optimally correct estimates of orthologs in polynomial time to obtain a tree-explainable orthology
graph. A natural question arises: how do best matches relate to orthology in the context of
networks? Addressing this question requires not only a deeper understanding of orthologs but also a
more detailed investigation of best matches and their relationship to orthology 
in network-based evolutionary scenarios.

Finally, our results open new avenues for further research in graph theory, particularly in
understanding the broader mathematical properties of orthology graphs and their relationships to
other graph classes or graph parameters. 
Investigating these connections could lead to novel insights that extend
beyond the scope of orthology inference.

\subsection*{Acknowledgements}
This work was funded by the German Research Foundation
  (DFG) (Proj.\ No.\ MI439/14-2). We thank the organizers of 
  the 40th TBI Winterseminar in Bled, Slovenia where this paper
  was finalized.

\bibliographystyle{spbasic}
\bibliography{lvll1}

\end{document}